\documentclass[%
 reprint,
superscriptaddress,
nofootinbib,
 amsmath,amssymb,
 aps,
prb,
]{revtex4-2}

\usepackage{physics}
\usepackage{xcolor}
\usepackage{graphicx}
\usepackage{bm}
\usepackage{hyperref}
\hypersetup{
    colorlinks=true,
    linkcolor=blue,
    urlcolor=cyan,
    citecolor=red,
    }
\usepackage[normalem]{ulem}
\usepackage{verbatim}
\usepackage{enumitem}

\begin{document}

\title{Dynamical purification and the emergence of quantum state designs from the projected ensemble} 

\author{Matteo Ippoliti}
\affiliation{Department of Physics, Stanford University, Stanford, CA 94305, USA}

\author{Wen Wei Ho}
\affiliation{Department of Physics, Stanford University, Stanford, CA 94305, USA}
\affiliation{Department of Physics, National University of Singapore, Singapore 117542}

\date{\today}

\begin{abstract}
Quantum thermalization in a many-body system is defined by the approach of local subsystems towards a universal form, describable as an ensemble of quantum states wherein observables acquire thermal expectation values.
Recently, it was demonstrated that the {\it distribution} of these quantum states can also exhibit universal statistics, upon associating each state with the outcome of a local projective measurement of the complementary subsystem. Specifically, this collection of pure quantum states -- called the projected ensemble -- can under certain conditions mimic the behavior of a maximally entropic, uniformly random ensemble, i.e., form a {\it quantum state-design}, representing a ``deeper'' form of quantum thermalization. In this work, we investigate the  dynamical process underlying this novel emergent universality. Leveraging a space-time duality mapping for one-dimensional quantum circuits, we argue that the physics of dynamical purification, which arises in the context of monitored quantum systems,
constrains the the projected ensemble's approach towards the uniform distribution. 
We prove that absence of dynamical purification in the space-time dual dynamics (a condition realized in dual-unitary quantum circuits with appropriate initial states and final measurement bases) generically yields exact state-designs for all moments $k$  at the same time, extending previous rigorous results [Ho and Choi, Phys. Rev. Lett. {\bf 128}, 060601 (2022)]. Conversely, we show that, departing from these conditions, dynamical purification can lead to a separation of timescales between the formation of a quantum state-design for moment $k=1$ (regular thermalization) and for high moments $k\gg 1$ (deep thermalization). Our results suggest that the projected ensemble can probe nuanced features of quantum dynamics inaccessible to regular thermalization, such as quantum information scrambling.
\end{abstract}

\maketitle

\tableofcontents

\section{Introduction \label{sec:intro} }

Generic isolated quantum many-body systems are expected to thermalize under their own dynamics. That is, such systems are believed to relax locally to a steady state given by a maximally entropic mixed state, up to constraints from global conservation laws~\cite{srednicki_chaos_1994, rigol_thermalization_2008, dalessio_quantum_2016}. 
This universal behavior arises because of the build-up of entanglement between a local subsystem and its complement, which serves as a `bath': upon ignoring the state of the latter, the former is captured by a reduced density matrix describing a statistical mixture of different configurations that has a universal form (the Gibbs ensemble).

\begin{figure*}
    \centering
    \includegraphics[width=\textwidth]{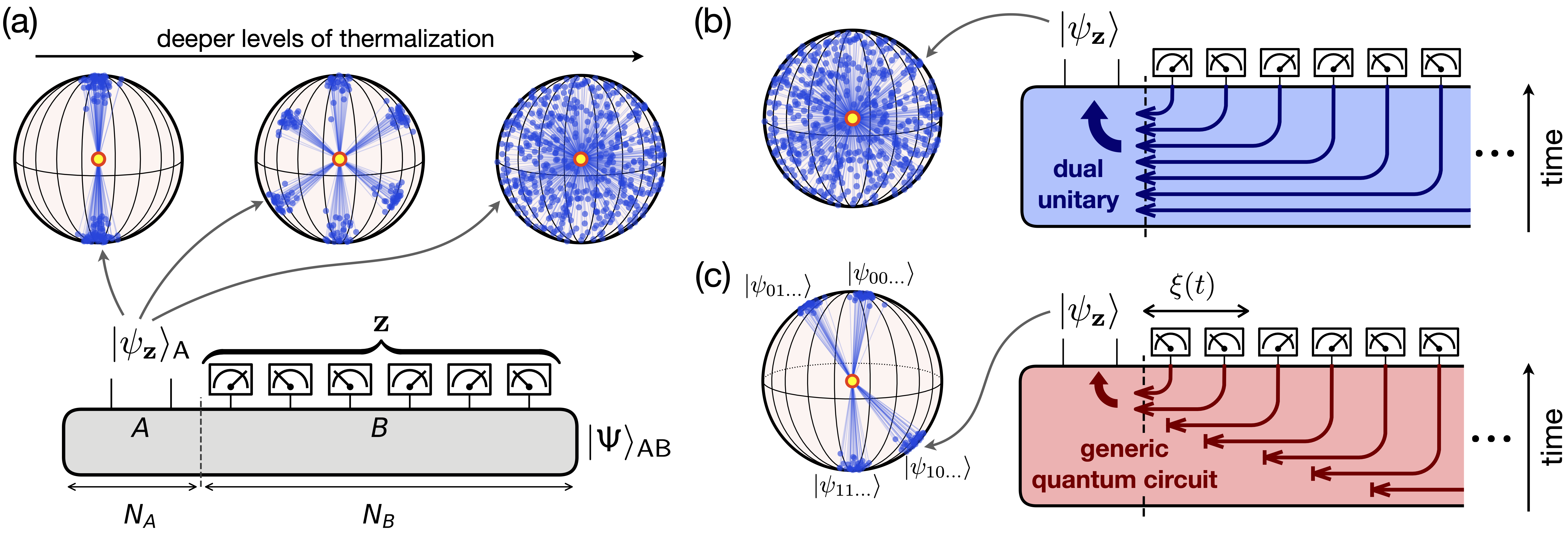}
    \caption{Main ideas of the paper. 
    (a) The projected ensemble. Bottom: a global many-body state $\ket{\Psi}$ on a composite system $AB$ is projectively measured in $B$ to yield bit-string outcome $\mathbf{z}$, and  a pure state $\ket{\psi_{\mathbf z}}$ on $A$, 
    with probability $p({\mathbf z})$.
    The collection $\{p({\mathbf z}), \ket{\psi_{\mathbf z}} \}$ constitutes the projected ensemble on $A$, and is a distribution over the Hilbert space. In this work, we are interested in projected ensembles generated from many-body states undergoing quantum chaotic dynamics.  
    Top: illustrations of different distributions that may emerge in the case ${\sf dim}(\mathcal{H}_A) = 2$ (Bloch sphere). In all cases, the weighted mean (yellow circle) of all the points is in the center of the Bloch sphere (regular thermalization at infinite temperature); however, the points can be more (right) or less (left) uniformly distributed, representing deeper forms of quantum thermalization. This can be quantified by the notion of quantum state-designs.
    (b, c) Schematic summary of our results. We focus on $(1+1)$-dimensional quantum circuits and study how  information from measurement outcomes gets propagated in space (represented by arrows from the  measurement location to   $A$), and their effect on the structure of the projected ensemble. 
    (b) In Sec.~\ref{sec:du} we identify a large class of quantum circuits where the propagation of information in space is perfect (i.e., unitary), called `dual-unitary' circuits, and show that typically 
    any two bitstrings $\mathbf z\neq \mathbf{z}'$ induce significantly distinct states $\ket{\psi_{\mathbf z}}$, $\ket{\psi_{\mathbf z'}}$, so that the states uniformly cover the Hilbert space (exact deep thermalization) in finite time. 
    (c) If the propagation of information in space is not perfect (generic quantum circuits), the influence of measurement outcomes decays over some finite length scale $\xi(t)$ (schematically represented by arrows terminating). As a result, the projected ensemble is highly degenerate, with $\ket{\psi_{\mathbf z}} \approx \ket{\psi_{\mathbf z'}}$ whenever bitstrings $\mathbf z$, $\mathbf{z}'$ agree in the first $\xi$ entries (the case of $\xi=2$ sketched). This limits the ability of the projected ensemble to uniformly cover the Hilbert space. In Section~\ref{sec:purification}, we identify the scaling of $\xi(t)$ and use it to bound the time taken to achieve deep thermalization.
    }
    \label{fig:outline}
\end{figure*}

Recently, a new perspective on  describing a local subsystem  of a quantum many-body system  was put forth by Ref.~\cite{choi_preparing_2023, cotler_emergent_2023}.
Instead of ignoring   the state of the bath, it is assumed that some knowledge of the bath's effect on the subsystem can be retained by the observer. 
Concretely, one can study a collection of {\it pure  states} of a local subsystem $A$ obtained by projectively measuring the complementary subsystem $B$ in a fixed, local basis, as sketched in Fig.~\ref{fig:outline}(a). 
These projected states,  together with their respective Born-rule probabilities, form the so-called {\it projected ensemble}, and can be thought of as  `unraveling' the reduced density matrix in terms of its constituent pure states  according to an observation of the bath. Notably, this goes beyond the standard framework of quantum thermalization, which depends solely on the reduced density matrix itself.

Intriguingly, Refs.~\cite{choi_preparing_2023, cotler_emergent_2023} found evidence of a novel universal behavior exhibited by the projected ensemble.
 In particular, it was observed through numerics and experiments that under quench dynamics of generic quantum-chaotic many-body systems without conservation laws or at infinite temperature, the distribution of projected states approaches that of a uniformly (i.e., unitarily-invariant, or Haar) random ensemble, independent of microscopic details, with an accuracy that increases with the bath size and the quench time. 
In quantum information theoretic language, the projected ensemble is said to approach a {\it quantum state-design}~\cite{renes_symmetric_2004, ambainis_quantum_2007}.
This behavior implies that the statistics of the local subsystem tends towards a maximally entropic distribution not just at the level of expectation values of local observables, but rather at the level of the Hilbert space. Thus, it can be viewed as a stronger notion of quantum thermalization, which can be dubbed ``deep thermalization''.

Subsequent work has established this novel phenomenon more firmly. An exactly-solvable instance of this behavior was provided in Ref.~\cite{ho_exact_2022}, which analyzed the non-integrable kicked Ising model (KIM) in certain parameter regimes and rigorously proved the emergence of {\it exact} quantum state-designs at finite time in quench dynamics.
The proof leveraged the so-called \emph{dual-unitarity} of the KIM---the property that its representation as a quantum circuit in $(1+1)$-dimensional space-time is unitary along both the time and space directions. 
There the projected states can be understood as arising from different unitary evolutions in the space direction, indexed by measurement outcomes on the bath; it was further shown that the unitary operators corresponding to the different measurement outcomes densely fill the unitary group (in the infinite-bath limit), giving rise to a quantum state-design.
Ref.~\cite{claeys_emergent_2022} further provided constructions of a class of quantum circuits that can be analyzed in similar fashion as the KIM, strongly suggesting the emergence of exact quantum state designs in these models too.

Despite such progress, much remains to be understood of this newly uncovered nonequilibrium universality.
An interesting question is to characterize in  finer detail the dynamical process of deep thermalization. 
To this end one can consider the time $t_k$ at which the projected ensemble's distribution approaches the uniformly-random one at the level of its $k$-th moment, that is, the time taken to form an (approximate) {\it quantum state $k$-design}~\cite{renes_symmetric_2004, ambainis_quantum_2007}.
For this reason we refer to $\{t_k\}$ as the ``design times''.
Note that as  the first moment ($k=1$) of the projected ensemble is precisely the reduced density matrix, $t_1$ is the time taken for regular thermalization to occur. 
The design times are by definition non-decreasing in $k$, and for the models numerically probed in Ref.~\cite{cotler_emergent_2023}, were generically found to increase with $k$.
On the other hand, Ref.~\cite{ho_exact_2022} found that for the KIM in the thermodynamic limit, remarkably, all design times   were finite and furthermore  coincided, being exactly equal to the subsystem size $N_A$. In other words, in this model, when regular thermalization (convergence of the $k=1$ moment)  occurs, deep thermalization (convergence of the $k > 1$ moments) occurs  concurrently. 
This immediately raises   questions about the origin of these different  behaviors, and more generally, what physics sets the separation in timescales between regular thermalization and deep thermalization. 

In this article, we make progress on these questions in two ways, focusing on dynamics in one spatial dimension.
Our approach is summarized in Fig.~\ref{fig:outline}(b,c). 
First, we prove that the dynamics of quantum circuits comprised of random local gates which are dual-unitary --- i.e., circuits which can be given an interpretation of inducing unitary dynamics in both space and time directions, just like the KIM --- with a suitable choice of compatible initial states and local measurement basis on the bath, {\it almost always}  have their projected ensembles become exactly Haar-distributed
at the thermalization time $t_1 = N_A$. This result extends the scope of the conclusions in Refs.~\cite{ho_exact_2022} (which focuses on the KIM) and \cite{claeys_emergent_2022},
and its derivation is the result of perfect transport of information across space in this class of circuits, Fig.~\ref{fig:outline}(b).
Second, we show that if the above assumptions (of dual-unitarity and compatible initial states/measurement basis) are violated, then higher design times $t_{k>1}$ can be separated from the regular thermalization time $t_1$. 
This separation is a consequence of the imperfect transport of information across space for generic dynamics, see Fig.~\ref{fig:outline}(c). Such loss of information can be understood in terms of \emph{dynamical purification}~\cite{gullans_dynamical_2020}, a phenomenon that arises in quantum systems subject to monitoring by an outside observer,
whereby an input mixed state gradually becomes pure over the course of the dynamics. 
Such monitored systems have been the focus of much attention lately due to the emergence of entanglement phases and phase transitions in their quantum trajectories \cite{skinner_measurement-induced_2019, li_quantum_2018, li_measurement-driven_2019, chan_unitary-projective_2019, gullans_dynamical_2020}.
Here we leverage insights from monitored dynamics to quantify the influence of distant measurement outcomes on the projected ensemble. We find that, due to dynamical purification, many states in the projected ensemble are essentially degenerate, which limits the ensemble's ability to uniformly cover the space. With increasing time in dynamics, this restriction is alleviated, and deep thermalization may be achieved.

This result is interesting as it reveals nuances in the different notions of local thermal equilibration in quantum many-body systems.
Regular thermalization, captured by the reduced density matrix, only probes the total build-up of entanglement between the subsystem and the bath, and is insensitive to how that entanglement is organized in space. In contrast, deep thermalization, captured by higher moments of the projected ensemble, is sensitive to the choice of local measurement basis, and thus to the structure of entanglement in space. 
This suggests a connection with quantum information scrambling~\cite{hayden_black_2007,hosur_chaos_2016, roberts_chaos_2017, nahum_operator_2018, von_keyserlingk_operator_2018, mi_information_2021}, the phenomenon by which local operators grow in size over time, spreading quantum information across the system.
Importantly, these phenomena can happen on different timescales. For example, chaotic systems feature an ``entanglement velocity'' $v_E$ governing the ballistic growth of entanglement (and thus thermalization)~\cite{kim_ballistic_2013, nahum_quantum_2017} and a ``butterfly velocity'' $v_B$ governing scrambling~\cite{shenker_black_2014,nahum_operator_2018, von_keyserlingk_operator_2018}. In this work we identify yet another quantity, the ``purification velocity'' $v_p$, bounding the formation of high designs. 
 
Beyond   fundamental theoretical interest,
  our findings have potential impact on practical applications too.
 Many important quantum information science protocols require the use of certifiably random states, such as in randomized tomography, cryptography or benchmarking~\cite{PhysRevA.77.012307, PhysRevLett.106.180504, PhysRevLett.116.170502,
UnforgeableQuantumEncryption, Bouland_2018, Haferkamp2020ClosingGap,
1663744,  Hayden_2004,  PhysRevX.6.041044, Kimmel_2017,  Arute_2019, huang_predicting_2020, 2021arXiv210104634A, PhysRevLett.126.190505}, which can be challenging to realize.
The fact that such desirable randomness can emerge naturally in generic chaotic dynamics with relatively modest control requirements (namely the ability to perform local measurements) promises a range of applications implementable in current-day quantum technologies.
Indeed, recent works have leveraged this phenomenon to propose new ways to benchmark~\cite{choi_preparing_2023} and perform state-learning via classical shadows, tailored for analog quantum simulators~\cite{tran_measuring_2022, mcginley_shadow_2022}. Our results on the separation of design times and the physics which affects them, give  us a way to quantitatively certify the randomness that emerges, and  may hence provide better understanding of the performance of such new quantum information science protocols.

Before proceeding, let us remark that Ref.~\cite{wilming_high-temperature_2022} already establishes a sharp connection between the onset of regular thermalization and the simultaneous formation of higher designs:
 whenever the reduced density matrix  on $A$ is close to being maximally mixed (i.e., at infinite temperature), then with high probability the projected ensemble forms an approximate quantum state-design, provided the measurement basis of the complementary subsystem is chosen at random from the Haar measure (which is   typically non-local and highly entangled). 
However, a setting which is arguably more natural for experiments is that of {\it spatially local} measurement bases, like the ones we consider in this work. Such bases are highly atypical (with respect to the Haar measure), and thus they can evade the result of Ref.~\cite{wilming_high-temperature_2022}, enabling nontrivial separations between the different design times.

The rest of the paper is organized as follows.
In Section~\ref{sec:setup} we review the projected ensemble formalism, as well as the concepts of space-time duality and dynamical purification in monitored systems which are important technical tools in this work.
Section~\ref{sec:du} contains our first main result, a proof of the emergence of exact state designs in random dual-unitary circuits.
In Section~\ref{sec:purification} we move away from dual-unitarity and present our second main result, an analytical connection between design formation and dynamical purification. 
Finally, we discuss our results and outline directions for future work in Sec.~\ref{sec:discussion}.


\section{Overview of relevant concepts \label{sec:setup} }

We begin by motivating and reviewing the theoretical framework behind the projected ensemble, Fig.~\ref{fig:outline}(a), first introduced  by Refs.~\cite{choi_preparing_2023, cotler_emergent_2023}. 
We also provide a high-level introduction to the concepts of space-time duality --- the idea of studying quantum circuits as evolution in the space, rather than time, direction, Fig.~\ref{fig:review}(a-c); and monitored quantum dynamics --- the dynamics of quantum systems evolving under the combined action of unitary transformations and measurements, Fig.~\ref{fig:review}(d), which will play central roles in our   analysis of the process of deep thermalization.


\subsection{The projected ensemble \label{sec:proj_ens} }

 \subsubsection{Motivation and definition}

Consider a quantum many-body system of spin-$1/2$ degrees of freedom (qubits)\footnote{Though we have specialized to qubit systems here, note that all discussions and results extend straightforwardly to the more general case of qudits or fermions. } on $N$ sites, described by a pure global wavefunction $|\Psi\rangle$. 
Suppose we are interested in describing the properties of a local subsystem $A$, comprised of spins $i=1,\cdots,N_A$. The conventional approach is to construct the reduced density matrix $\rho_A$ by tracing out the complementary subsystem $B$ (intuitively, the `bath'):
\begin{align}
    \rho_A = \Tr_B \left( |\Psi\rangle \langle \Psi| \right),
\end{align}
in which expectation values of observables $O$  supported   on region $A$ can   be computed    via
\begin{align}
    \langle O \rangle = \text{Tr}\left( \rho_A O \right).
    \label{eqn:standard_observable}
\end{align}
If $|\Psi\rangle = |\Psi(t)\rangle$ is a state obtained in dynamics, then   studying how expectation values of local observables settle to an equilibrium value, $\langle O(t)\rangle \to \langle O \rangle_\text{eq.}$, is equivalent to studying how the reduced density matrix relaxes to an equilibrium ensemble, $\rho_A(t) \to \rho_\text{eq.}$. According to the general statistical-mechanical principle of   maximization of entropy, we expect $\rho_\text{eq.}$ to be    given by a Gibbs state parameterized by Lagrange multipliers corresponding to different globally conserved quantities (modulo ergodicity-breaking scenarios  like many-body localization~\cite{nandkishore_many-body_2015, abanin_colloquium_2019}). In particular, with only energy conservation, this takes the form
\begin{align}
    \rho_\text{eq.} \propto \Tr_B  \left( e^{-\beta H} \right),
    \label{eqn:Gibbs}
\end{align}
where $\beta$ is the inverse temperature set by the conserved energy of the initial state. We refer to such an equilibriation scenario as ``regular thermalization''.

The  formalism of a reduced density matrix as a description of local properties of a subsystem is complete, but importantly   under the assumption that knowledge of the complementary subsystem $B$ is inaccessible (or lost) to the observer. However, in a new generation of experimental systems---quantum simulators~\cite{altman_quantum_2021}---this assumption need not always hold. 
For example,  in systems like cold atoms in optical lattices with quantum gas microscopes, individually trapped Rydberg atoms or ions, and superconducting circuits, microscopic read-out of the entire system is routinely performed, whereupon global {\it bit-strings} $\mathbf Z \in \{0,1\}^N$ are collected.
By splitting such a bit-string into its restrictions to subsystems $A$ and $B$, $\mathbf{Z} = (\mathbf{z}_A,\mathbf{z}_B)$ with $\mathbf{z}_A\in\{0,1\}^{N_A}$ and $\mathbf{z}_B\in\{0,1\}^{N_B}$ (here $N=N_A+N_B$), we gain a joint classical snapshot of the states of both the subsystem of interest and the bath; in other words, {\it correlations} between the subsystem and bath are directly accessible. 
  Explicitly, statistics of the subsystem $A$ can be studied  {\it conditioned} upon observing   state  $\mathbf{z}_B$ of the bath $B$. Such information cannot be captured solely by the reduced density matrix and necessitates the development of a new theoretical framework.

The {\it projected ensemble} formalism  precisely achieves this goal. 
We consider the case of projective measurements performed in the computational basis---which, as we argued above, is of practical relevance to quantum simulators---but we note this assumption can be relaxed to allow for other choices too (as we will do later, and as was also considered in \cite{claeys_emergent_2022, wilming_high-temperature_2022}). 
Concretely, a measurement outcome on $B$ will be labeled by a bit-string $\mathbf z \in \{0,1\}^{N_B}$ (we drop the subscript  in $\mathbf{z}_B$ for convenience), so that following the measurement, the global wavefunction is updated according to the Born rule as
\begin{align}
    |\Psi\rangle \mapsto \left( \mathbb{I}_A \otimes |\mathbf z\rangle \langle \mathbf z|_B \right) |\Psi\rangle/\sqrt{p({\mathbf z})},
    \label{eqn:Bornupdate}
\end{align}
which occurs with probability
\begin{align}
    p({\mathbf z}) = \langle \Psi| \left( \mathbb{I}_A \otimes |\mathbf z\rangle \langle \mathbf z|_B \right) |\Psi\rangle.
    \label{eqn:Bornprobabilities}
\end{align}
Consequently, the state on $A$ will be in a {\it pure} state indexed by the measurement outcome $\mathbf z$,
\begin{align}
    |\psi_{\mathbf z}\rangle = \left( \mathbb{I}_A  \otimes \langle \mathbf z|_B \right) |\Psi\rangle/\sqrt{p({\mathbf z})}.
    \label{eqn:BornA}
\end{align}
The projected ensemble is then defined to be the set of such pure states and their respective probabilities indexed by $\mathbf{z}\in\{0,1\}^{N_B}$:
\begin{align}
    \mathcal{E} := \{ p(\mathbf{z}), |\psi_{\mathbf z}\rangle \},
\end{align}
which is a { probability distribution} on the Hilbert space\footnote{Such distributions are also known as {\it geometric quantum states}~\cite{anza_beyond_2021, anza_quantum_2022}.}  $\mathcal{H}_A$ of $A$.

\subsubsection{Moments of the projected ensemble and their information content} 

A probability distribution may be characterized by its moments.
For the projected ensemble, the $k$-th moment is captured by the object
\begin{align}
    \rho^{(k)} = \sum_{\mathbf z} p(\mathbf{z}) \left( |\psi_{\mathbf z} \rangle \langle \psi_{\mathbf z} | \right)^{\otimes k},
\end{align}
which is a density matrix defined on $k$  replicas  of the Hilbert space $\mathcal{H}_A$.
It can be readily verified that the mean, $k=1$, is the reduced density matrix $\rho^{(1)} = \rho_A$. An example of information captured by the projected ensemble but not by the reduced density matrix is the quantity
\begin{align}
    \mathbb{E}_{\mathcal{E}}\left[ \langle O\rangle_{\mathbf z}^2 \right] \equiv \sum_{\mathbf z} p({\mathbf z}) \left( \langle \psi_{\mathbf z} | O |\psi_{\mathbf z} \rangle \right)^2,
    \label{eqn:Obs}
\end{align}
which is related to the ensemble variance of {\it conditional} expectation values $\langle O\rangle_{\mathbf z} = \langle \psi_{\mathbf z} |O|\psi_{\mathbf z}\rangle$---the expectation of an operator $O$ on $A$ conditioned upon observing the bath in (classical) state $\ket{\mathbf{z}}_B$. 
This quantity can in fact be expressed as the expected value of a ``higher-order'' observable evaluated in a specific density matrix on a $k$-fold replicated space, namely $\mathbb{E}_{\mathcal{E}}\left[ \langle O\rangle_{\mathbf z}^k \right] = \Tr\left( \rho^{(k)} O^{\otimes k}  \right)$, with $k = 2$ (compare this expression to  Eq.~\eqref{eqn:standard_observable} for a ``regular'' observable).
The quantity in Eq.~\eqref{eqn:Obs} and its generalizations to higher $k$ have been measured in a Rydberg-atom-based quantum simulator~\cite{choi_preparing_2023}, specifically with the choice $O = \ket{\mathbf {s}} \bra{ \mathbf{s}}_A$, a projector onto bit-string $\mathbf {s} \in \{0,1\}^{N_A}$.
This yields information on the probability of observing bit-string $\mathbf {s}$ on $A$, conditioned upon having observed bit-string $\mathbf z \in \{0,1\}^{N_B}$ on the bath.

\subsubsection{Deep thermalization, quantum state-designs, and design times}
\label{sec:design_times}

Given that it is believed that the first moment of the projected ensemble $\rho^{(1)} = \rho_A$   tends to a universal equilibrium state $\rho_\text{eq.} \propto \Tr_B\left(e^{-\beta H} \right)$ in generic quantum many-body dynamics, an immediate question that arises is if higher moments $\rho^{(k)}$ similarly equilibrate to universal ensembles. In general, the form of such ensembles is not fully known (see however Ref.~\cite{goldstein_distribution_2006, goldstein_universal_2016}), but there is a particularly clean limiting scenario that can be considered: the case of systems without symmetries or conservation laws, in which case the principle of maximization of entropy suggests the unitarily-invariant (Haar) ensemble, whose $k$th moment is given by
\begin{align}
    \rho_H^{(k)} = \int_{\psi \sim \text{Haar}(\mathcal{H}_A)} d\psi \left( |\psi\rangle \langle \psi | \right)^{\otimes k}
    = \frac{\Pi_\text{symm}}{\binom{2^{N_A}+k-1}{k}},
    \label{eqn:Haar_ensemble}
\end{align}
where $\Pi_\text{symm}$ is a projector on the symmetric sector of the replicated Hilbert space $\mathcal{H}_A^{\otimes k}$.
It is natural to conjecture that in dynamics without explicit conservation laws (like time-periodic   systems or quantum circuit dynamics), or in systems at infinite temperature, $\rho^{(k)}$ equilibrates to such a maximally-entropic ensemble, i.e.,
\begin{align}
    \rho^{(k)}(t) \xrightarrow{t\to\infty} \rho^{(k)}_H.
\end{align}
For $k=1$, this reproduces Eq.~\eqref{eqn:Gibbs} with $\beta = 0$.

Mathematically,   convergence to the uniform ensemble can be quantified   by the normalized distance
 \begin{align}
     \Delta^{(k)}_\alpha \equiv \frac{\| \rho^{(k)} - \rho^{(k)}_H \|_\alpha}{\| \rho_H^{(k)} \|_\alpha},
     \label{eq:delta_def}
 \end{align}
where $\|\cdot\|_\alpha$ is the Schatten norm of index $\alpha$.
In quantum information theoretic language,  when $\Delta^{(k)}_\alpha=0$ the ensemble $\mathcal{E}$ of pure states is said to form a {\it quantum state $k$-design}, as it is reproducing the $k$th moment of the Haar random ensemble,  Eq.~\eqref{eqn:Haar_ensemble}~\cite{renes_symmetric_2004, ambainis_quantum_2007}.  
 A  property that may be desired of these distances is that they obey \emph{monotonicity}, i.e., that $\Delta^{(k+1)}_\alpha \geq \Delta^{(k)}_\alpha$ for any $k$. 
This is desirable because it implies that, if $\mathcal{E}$ is an $\epsilon$-\emph{approximate} quantum state $k$-design (that is,   $\Delta^{(k)}_\alpha \leq \epsilon$), then it is also an $\epsilon$-approximate quantum state $k'$-design for all $k' \leq k$.
 In Appendix~\ref{app:monotonicity} we show that the distances defined with the indices $\alpha \geq 1$ all obey such a property \cite{Daniel}. Note that the commonly-used indices include $\alpha = 1$ (trace norm), $\alpha = 2$ (Frobenius norm), and $\alpha = \infty$ (operator norm).

The measures defined in Eq.~\eqref{eq:delta_def} to quantify closeness to a quantum state $k$-design include one which will be easier to work with analytically, which
is based on the so-called {\it frame potential} of the ensemble $\mathcal{E}$~\cite{roberts_chaos_2017, renes_symmetric_2004}, defined for each $k$ as
\begin{align}
    F^{(k)} \equiv \sum_{\mathbf{z}, \tilde{\mathbf z}} p_{\mathbf z}p_{\tilde{ \mathbf z}} |\langle \psi_{\mathbf z} | \psi_{\tilde{\mathbf z}} \rangle |^{2k} = \Tr\left[ \left(\rho^{(k)} \right)^2 \right].
\end{align}
This is nothing more than the purity of the density matrix $\rho^{(k)}$. 
Indeed, one can see that for $\alpha = 2$, Eq.~\eqref{eq:delta_def} yields
\begin{equation}
\left( \Delta_2^{(k)} \right)^2 = \frac{F^{(k)}}{F_H^{(k)}} - 1,
\end{equation}
where $F_H^{(k)} \equiv \binom{2^{N_A}+k-1}{k}^{-1}$ is the frame potential of the Haar ensemble.
It follows that the frame potential obeys the bound
\begin{align}
    F^{(k)} \geq  \binom{2^{N_A}+k-1}{k}^{-1} \label{eq:haar_fp}
\end{align}
with equality if and only if $\mathcal{E}$ forms an exact quantum state $k$-design, $\rho^{(k)} = \rho^{(k)}_H$. 

Given the above measures of closeness to a uniformly-random ensemble, one may ask about the time taken for the projected ensemble to achieve a given state $k$-design within some accuracy $\epsilon$ over the course of quench dynamics. 
To this end we can define for each $k \geq 1$, and for each choice of Schatten-index $\alpha$, a {\it design time} $t_{k,\alpha}$ as
\begin{align}
    t_{k, \alpha} = \min_t \left( \Delta^{(k)}_\alpha (t) < \epsilon \right), \label{eq:approx_design_time}
\end{align}
which is the minimum time such that the chosen distance to the uniformly-random ensemble falls below   some arbitrarily  small threshold $\epsilon > 0$.
This definition naturally encompasses the thermalization time, $t_1$, but also an infinite sequence of higher design times $t_{k>1}$, which may capture features of the dynamics beyond regular thermalization, that we dub {\it deep} thermalization. 
Note that from monotonicity of the normalized distances $\Delta^{(k)}_\alpha$ for $\alpha \geq 1$ (Appendix~\ref{app:monotonicity}), the design times $t_{k,\alpha}$ for such indices are also monotonic: $t_{k+1,\alpha} \geq t_{k,\alpha}$.
Past works on approximate quantum state designs have used $\alpha = 1$~\cite{cotler_emergent_2023, choi_preparing_2023} and $\alpha = \infty$~\cite{ambainis_quantum_2007, low_pseudo-randomness_2010}; in this work it will be technically more convenient to use $\alpha = 2$, and we henceforth drop the subscript $\alpha$ with the understanding that we refer to $\alpha = 2$. 
While we do not expect the choice of $\alpha$ to qualitatively affect the scaling of design times, it would be interesting to investigate this dependence in future work.

In the rest of this paper, we focus on quench dynamics generated by quantum circuits in one-dimension with gates arranged in a brickwork pattern, which serve as toy models for quantum many-body dynamics that preserve only the important minimal ingredients of locality and unitary, and study the behavior of design times of their projected ensemble.


\subsection{Space-time duality and dual-unitarity \label{sec:setup_std}}
\begin{figure*}
    \centering
    \includegraphics[width=\textwidth]{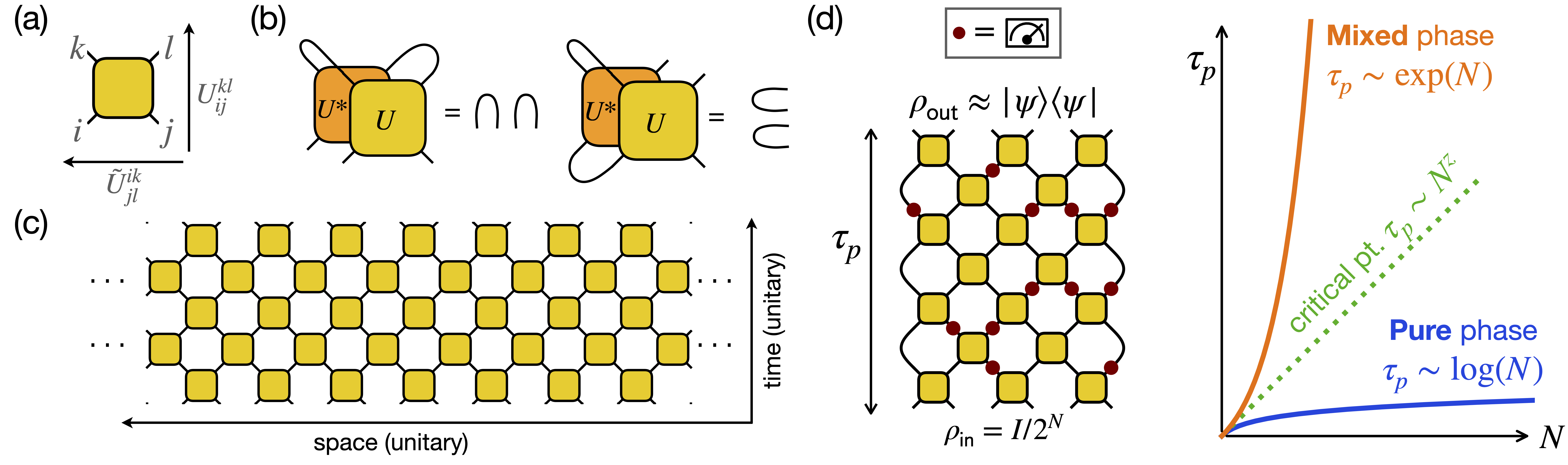}
    \caption{Review of space-time duality and dynamical purification.
    (a) Space-time duality: a two-qubit unitary gate $U$ may be viewed, by a different grouping of tensor indices, as a linear operator $\tilde{U}$ acting in space (right to left in our convention). $\tilde{U}$ is generically non-unitary.
    (b) Dual-unitarity: a gate $U$ is dual-unitary (DU) if it obeys $\tilde{U}\tilde{U}^\dagger = \mathbb{I}$ as well as $UU^\dagger = \mathbb{I}$, corresponding to unitary operations in both space and time.
    (c) Stacking DU gates in a brickwork pattern yields a DU circuit, unitary along both space and time.
    (d) Dynamical purification. A fully-mixed state on $N$ qubits $\rho_{\rm in} = \mathbb{I}/2^N$ is evolved under a monitored circuit, featuring both unitary gates and non-unitary measurements. As a result of measurements, the state eventually becomes pure, $\rho_{\rm out} \approx \ketbra{\psi}{\psi}$, over a time scale $\tau_p$ ({\it purification time}). As a function of the strength or density of measurements, distinct {\it purification phases} may arise, sharply distinguished by the asymptotic scaling of $\tau_p$ vs the number of qubits $N$. In this work we make use of the scaling $\tau_p \sim \exp(N)$ in the {\it mixed phase}.
    }
    \label{fig:review}
\end{figure*}

Recently, significant progress in our understanding of the projected ensemble and the formation of state designs \cite{ho_exact_2022, claeys_emergent_2022} has been achieved by leveraging the concept of {\it space-time duality}. 
This is a transformation of one-dimensional quantum circuits that exchanges the roles of space and time in the dynamics~\cite{banuls_matrix_2009, hastings_connecting_2015, lerose_influence_2021, garratt_local_2021, ippoliti_postselection-free_2021, ippoliti_fractal_2022, lu_spacetime_2021, garratt_many-body_2021} (recently extended also to higher dimensional circuits~\cite{lu_spacetime_2021, jonay_triunitary_2021} as well as continuum field theories~\cite{bertini_growth_2022}). 
At the microscopic level, the transformation acts on individual two-qubit unitary gates $U$ that make up a circuit, which has matrix elements $U_{ij}^{kl} = \bra{k}\otimes\bra{l} U \ket{i}\otimes\ket{j}$, where $i,j,k,l \in \{0,1\}$ index the input and output states of qubits in the computational basis.
The idea, illustrated in Fig.~\ref{fig:review}(a), is to  view the action of the gate, by definition an evolution operator in time, as instead an `evolution' operator {\it in space},
that is, consider the right legs $j,l$ of the tensor as inputs and the left legs $i,k$ as outputs. Doing so defines a different matrix, $\tilde{U}$, whose entries are a permutation of those of $U$, namely\footnote{
This corresponds formally to $\tilde{U} = \textsf{SWAP} \cdot (U\cdot \textsf{SWAP})^{T_2}$, where ${\sf SWAP} = e^{i \frac{\pi}{4} \boldsymbol{\sigma}_1 \cdot \boldsymbol{\sigma}_2}$ exchanges the state of the two qubits it acts on, $\textsf{SWAP}_{ij}^{kl} = \delta_{il}\delta_{jk}$, and $T_2$ denotes the partial transpose acting on the second qubit, $(A^{T_2})_{ij}^{kl} = A_{il}^{kj}$. Note other conventions exist but are all related by unitary operations ($\textsf{SWAP}$ gates).}
$ \tilde{U}_{jl}^{ik} = U_{ij}^{kl}$.
Informally, one can think of this transformation as simply rotating the tensor network diagram of $U$ in Fig.~\ref{fig:review} by 90${}^\text{o}$ clockwise, which defines a new tensor.

This transformation generically breaks unitarity---i.e., the space-time dual operator $\tilde{U}$ is typically {\it not} itself unitary.
However, there exists a set of gates  which have the property of being \emph{dual-unitary}  (DU)~\cite{akila_particle-time_2016, bertini_exact_2018, gopalakrishnan_unitary_2019, claeys_maximum_2020}, that is, {\it both} $U$ and its space-time dual $\tilde{U}$ are unitary. This means they obey the diagrammatic relations constraining different tensorial contractions shown in Fig.~\ref{fig:review}(b). A simple example is the $\textsf{SWAP}$ gate: it can be easily seen  that $\widetilde{\textsf{SWAP}} = \textsf{SWAP}$, making the gate DU, and in particular self-dual.
More generally, DU gates on two qubits\footnote{Note one can straightforwardly also define space-time duality and dual-unitary gates on qudits of general dimension~\cite{rather_creating_2020, gutkin_local_2020, claeys_ergodic_2021}; here we focus on qubits for concreteness.} form a submanifold of co-dimension $2$ in the unitary group $U(4)$ on two qubits. Neglecting an overall phase, these can be parametrized as~\cite{bertini_exact_2019} 
\begin{equation}
\mathfrak{DU} = \{(r_1 \otimes s_2) \mathsf{SWAP} e^{-iJZ_1Z_2} (u_1 \otimes v_2)\}
\label{eqn:DU_qubits}
\end{equation}
where $r,s,u,v\in SU(2)$, $J\in [0,\pi/4]$ (giving a total of 13 parameters), and $Z$ the standard Pauli-z operator. 

Brickwork circuits built out of compositions of such DU gates, called dual-unitary circuits~[Fig.~\ref{fig:review}(c)], which we shall refer in the rest of the paper as `DU' as well, are thus unitary in both the space and time directions. 
This condition gives a useful interpretation of a system of $N$ qubits evolving for $t$ time-steps by a DU quantum circuit, as being equivalent to $t$ qubits evolving for $N$ `time'-steps by another (possibly different, but still unitary) quantum circuit.
Indeed, dual-unitarity accords a great deal of analytical control in the study of quantum dynamics, and has enabled much  progress on questions ranging from the emergence of random matrix theory in spectral statistics, to the growth of entanglement and the decay of correlation functions in strongly-interacting systems~\cite{bertini_exact_2018, bertini_entanglement_2019, bertini_exact_2019, gopalakrishnan_unitary_2019, claeys_maximum_2020, bertini_scrambling_2020, kos_correlations_2021, piroli_exact_2020, claeys_ergodic_2021, bertini_random_2021}. 

\begin{figure}
    \centering
    \includegraphics[width=\columnwidth]{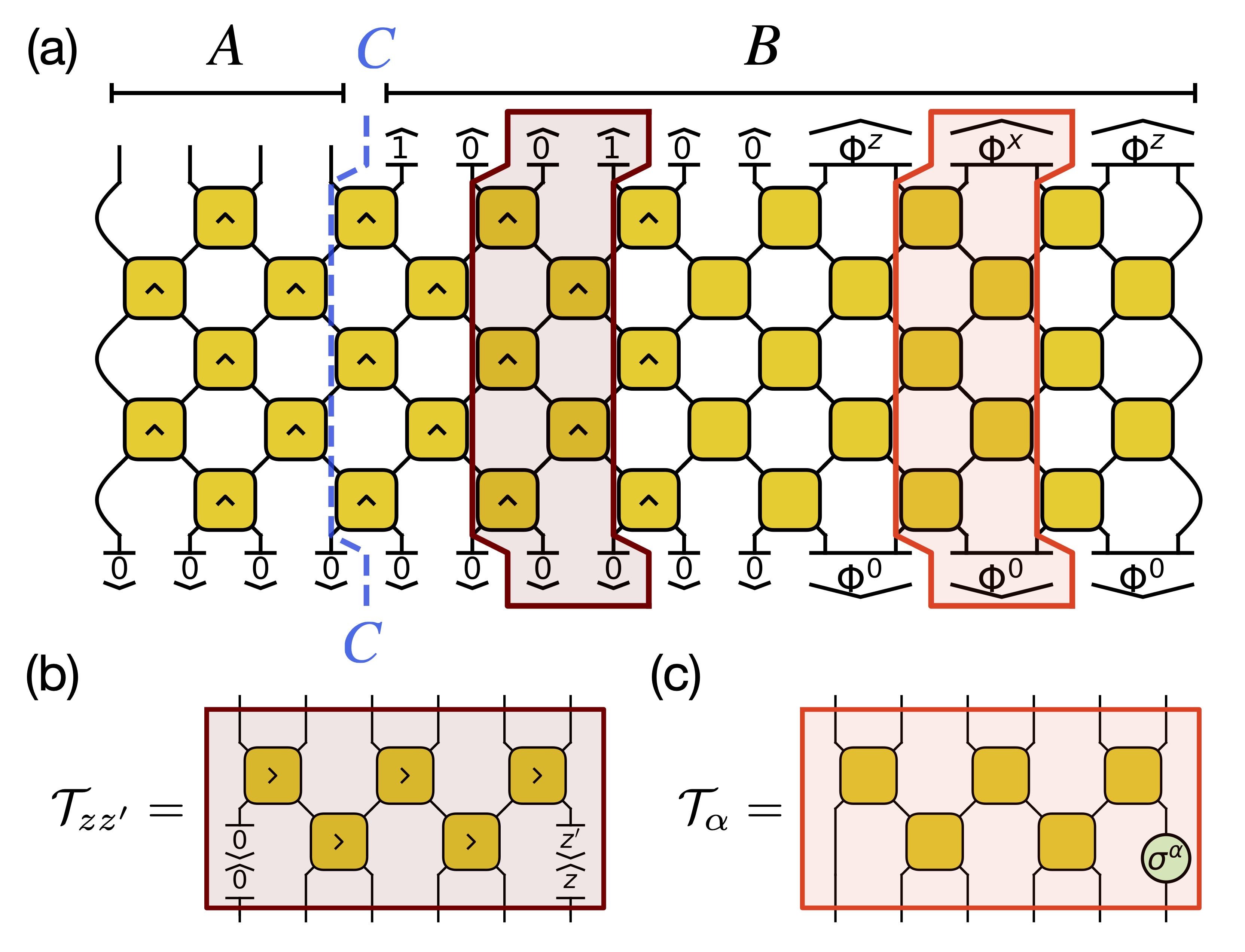}
    \caption{
    (a) Setup of the projected ensemble on a one-dimensional brickwork unitary circuit. 
    Yellow squares are two-qubit unitary gates. The caret points in the direction of unitarity; gates without a caret are DU.
    The initial state is a local product state either in the computational basis ($\ket{0}$) or in the Bell basis ($\ket{\Phi^0} \propto \ket{00} + \ket{11}$). 
    The final state is projectively measured in $B$, either in the computational basis (yielding outcomes $z_i \in \{0,1\}$) or in the Bell basis (yielding outcomes $\alpha_i \in \{0,x,y,z\}$). 
    A pure state in subsystem $A$ (dangling bonds) remains after the measurements. 
    We also display the time-like subsystem $C$ (dashed line) at the boundary between $A$ and $B$. 
    In this example, $N_A=4$, $N_B=12$, and $t=5$.
    (b) Transfer matrix $\mathcal{T}_{zz'}$ corresponding to the darker shaded region in (a). The initial state and final measurement map onto projectors $\ketbra{0}{0}$ and $\ketbra{z}{z'}$ visible at the edges of $\mathcal{T}_{zz'}$. This transfer matrix is not unitary in general.
    (c) Transfer matrix $\mathcal{T}_{\alpha}$ corresponding to the lighter shaded region in (a). The initial state and final measurement map onto unitary single-qubit operations ($\mathbb{I}$ and $\sigma^\alpha$ at the left and right edge, respectively). This transfer matrix is unitary. 
    }
    \label{fig:transfer_matrix}
\end{figure}

Dual-unitarity has provided a handle to rigorously understand deep thermalization, or the formation of exact state designs in dynamics,   as well. Specifically, Ref.~\cite{ho_exact_2022} considered the quantum circuit corresponding to the dynamics of the kicked Ising model (KIM) with couplings tuned to a special point, which yields a notable example of a DU circuit
(in Appendix \ref{app:KIM} we elaborate on the model and explain its dual-unitary nature).
The key idea elucidated in that work is to invoke space-time duality, in order to view the projected ensemble as being generated by a set of quantum evolutions in space, with particular measurement outcomes on the bath determining the particular evolution. 
This is sketched in Fig.~\ref{fig:transfer_matrix}(a) for the general case of dynamics under a one-dimensional brickwork circuit: one sees that a   projected state on subsystem $A$ can be thought of as first arising from evolution  of an initial state defined on the right boundary (which has size equal to the depth of the original circuit evolution), through a sequence of {\it transfer matrices} $\mathcal{T}_\alpha$ (purple/red boxes in Fig.~\ref{fig:transfer_matrix}(a)) applied from right to left, each of which is defined via space-time-dual operators $\tilde{U}$ and projections onto the initial states, as well as projections onto   measurement basis states according to measurement outcome $\alpha$ on the bath $B$. This yields a state that lives on the time-like cut $C$ (dashed blue line in Fig.~\ref{fig:transfer_matrix}(a)), which is finally mapped to spatial region $A$ by $W$, defined to be the part of the circuit that lives to the left of the dashed line.

Ref.~\cite{ho_exact_2022} considered dynamics under the KIM with couplings tuned to a special point and certain initial states and measurement bases, which together give rise to {\it unitary} transfer matrices $\mathcal{T}_\alpha$ (see Appendix~\ref{app:KIM}).
Thus, the projected ensemble can be understood as arising from a collection of random  unitary quantum circuits. 
Further, they  showed that the set $\{ \mathcal{T}_\alpha \}$ constitute a {\it universal ``gate'' set} in quantum computation --- that is, it is possible to approximate any desired unitary acting on $C$ arbitrarily well with some long concatenation of $T_\alpha$ operators. 
In the context of the projected ensemble, this implies that 
in the limit of an infinitely large measured bath $B$, the distribution of randomly-evolved states on the time-like cut $C$ is {\it uniform} over the Hilbert space. Lastly, the map $W$ from $C$ to $A$, under the same conditions that guaranteed unitarity of $\mathcal{T}_\alpha$, can also be seen to be an isometry ($WW^\dagger = \mathbb{I}_A$), provided $|A|\leq |C|$ (that is, provided the evolution time is long enough). 
Then, the desired conclusion (that the projected ensemble at $A$ is Haar-randomly distributed) follows from the mathematical fact that the projection of a Haar-random state from a higher-dimensional Hilbert space $\mathcal{H}_C$ to a lower-dimensional one $\mathcal{H}_A$ is still Haar-random on the smaller space $\mathcal{H}_A$.

The above reasoning is rather general, and not exclusively tied to the particular example of the KIM. 
Thus, it is reasonable to expect that a similar result ought to hold also in other quantum circuits possessing similar properties as the KIM---namely, perfect transport of information in the spatial direction.
This naturally leads us to consider dynamics under general DU circuits.
However, close inspection of the logic underlying the KIM result immediately yields that dual-unitarity of the gates alone is not sufficient to guarantee unitarity of the transfer matrices $\mathcal{T}$, a key step of the proof: this is because the initial states and final measurements, which entail projections onto particular basis states, can (and in fact, in general) map under space-time duality to non-unitary operations at the edges of $\mathcal{T}$.
An example of a non-unitary transfer matrix is shown in Fig.~\ref{fig:transfer_matrix}(b), $\mathcal{T}_{zz'}$ labelled by computational basis measurement outcomes $z, z'\in \{0,1\}$.
It is however possible, in analogy with the KIM case, to find special initial states and measurement bases for a DU circuit that ensure unitarity of $\mathcal{T}$  \cite{claeys_emergent_2022}; this defines the conditions ``DU${}^+$''\footnote{Note Ref.~\cite{claeys_emergent_2022} introduces a related notion of ``solvable measurement scheme'' based on the spectrum of the transfer matrix; our notion of DU${}^+$ implies that notion.}.
A prime example of a DU${}^+$ circuit is shown in Fig.~\ref{fig:transfer_matrix}(c):  quantum evolution consisting of DU gates, initial states of nearest-neighbor Bell pairs $\ket{\Phi^0} = (\ket{00} + \ket{11})/\sqrt{2}$, and final measurements in the Bell-pair basis $\{ \ket{\Phi^\alpha} = (\sigma^\alpha\otimes I)\ket{\Phi^0}: \alpha = 0,x,y,z\}$. 
One can see this yields a unitary transfer matrix $\mathcal{T}_\alpha$ in the space direction\footnote{This follows from the fact that $\braket{ij}{\Phi^\alpha} = (\sigma^\alpha)_{ij}$, where $i,j$ label states in the computational basis; see also Fig.~\ref{fig:du}(a).}~\cite{piroli_exact_2020, claeys_emergent_2022}.

Of course, the subtle point that needs to be addressed in order to generalize the KIM result, is to prove that the space-time dual dynamics covers the unitary group, or equivalently show that the unitary transfer matrices $\mathcal{T}$ form a universal gate set. In Sec.~\ref{sec:du}, we  will prove that this is almost always the case, for DU${}^+$ circuits which are spatiotemporally-random.


\subsection{Monitored dynamics and dynamical purification \label{sec:setup_monitored}}

In this work, we are also interested in the behavior of the projected ensemble under general quantum circuit evolution, i.e., such that   DU${}^+$ conditions are violated (e.g., by choosing gates outside $\mathfrak{DU}$ and/or incompatible initial states and measurement bases).
In this scenario, while a projected state  may still be viewed as arising from a concatenation of transfer matrices $\mathcal{T}$, generically these matrices become non-unitary, see Fig.~\ref{fig:transfer_matrix}(b).
Thus, the proof employed in Ref.~\cite{ho_exact_2022} to derive exact state designs  ceases to apply, and it becomes possible to have a nontrivial separation between distinct design times, i.e., $t_1 < t_k$ for $k>1$ (where the $t_k$'s are now defined in the $\epsilon$-approximate sense, as in Eq.~\eqref{eq:approx_design_time}). 

Usefully, evolution by such a sequence of non-unitary transfer matrices $\mathcal{T}$ can be   interpreted as quantum trajectories in an incarnation of \emph{monitored quantum dynamics}~\cite{ippoliti_postselection-free_2021, ippoliti_fractal_2022, lu_spacetime_2021}, a topic that has received significant attention in recent years, as it represents a new paradigm for nonequilibrium phase structure~\cite{skinner_measurement-induced_2019, li_quantum_2018, li_measurement-driven_2019, chan_unitary-projective_2019, gullans_dynamical_2020, gullans_scalable_2020, bao_theory_2020, choi_quantum_2020, jian_measurement-induced_2020, ippoliti_entanglement_2021, lavasani_measurement-induced_2021, fidkowski_how_2021, fan_self-organized_2021, lavasani_topological_2021, szyniszewski_universality_2020, nahum_entanglement_2020, nahum_measurement_2021, li_statistical_2021, li_entanglement_2021, agrawal_entanglement_2022, potter_entanglement_2022}. 
This subject deals with quantum evolutions  where unitary gates coexist with postselected (weak or projective) measurements.
To see why monitored dynamics may become relevant to the projected ensemble, consider the two distinct reasons for departure from the DU${}^+$ conditions:
\begin{itemize}
    \item[(i)] {\bf Non-DU gates}. Given a unitary $U$, its space-time dual operator $\tilde{U}$ is generically \emph{not} unitary, and it can always be decomposed as $\tilde{U} = 2MV$, where $V$ is unitary and $M$ is a positive matrix that describes a quantum measurement on two qubits\footnote{Formally, it is an entry in a POVM set, e.g., $\{M, \sqrt{\mathbb{I}-M^\dagger M}\}$. Note the choice of normalization ensures $0\leq M^\dagger M\leq \mathbb{I}$.} with a fixed outcome. 
    As an example, consider $U = \mathbb{I}$: employing a  space-time duality transformation, defined above,
    we have\footnote{This follows from noting that $\braket{ij}{\Phi^0} = \delta_{ij}/\sqrt{2}$, so that $\bra{ij} \mathbb{I} \ket{kl} = \delta_{ik}\delta_{jl} = 2\braket{ik}{\Phi^0} \braket{\Phi^0}{jl}$.}   $\tilde{U} = 2\ket{\Phi^0} \bra{\Phi^0}$, i.e., $V = \mathbb{I}$ and $M$ is a projective measurement in the Bell basis $\{\ket{\Phi^\alpha}\}$, with forced outcome $\alpha = 0$.
    (Note that if $U\in \mathfrak{DU}$ then the measurement is trivial, $M\propto \mathbb{I}$.)
    \item[(ii)] {\bf Incompatible initial state and final measurement basis}. In Fig.~\ref{fig:transfer_matrix}(a) we show two choices of bases: the computational basis $|z\rangle$ and   the Bell basis $|\Phi^\alpha\rangle$ for pairs of neighboring qubits. As can be seen diagrammatically in Fig.~\ref{fig:transfer_matrix}(b), the former dualize to projective measurements on the edge qubits. The latter instead dualize to unitary operations, as shown in Fig.~\ref{fig:transfer_matrix}(c). Different choices besides the computational or Bell basis may lead to intermediate results, i.e., weak measurements.
\end{itemize}

To briefly review monitored dynamics and concepts within relevant for our present work, consider for simplicity the time evolution of a fully-mixed initial state $\rho(0)  = \mathbb{I}/2^N$  under entangling unitary gates, as well as local measurements performed on randomly selected sites in space with a certain probability, or rate; see Fig.~\ref{fig:review}(d).
In an instance of dynamics up to time $t$, the system   evolves into a state $\rho_{\mathbf m}(t)$, where $\mathbf{m}$ labels the (classical) measurement record extracted. Repeating the dynamics  yields a random measurement outcome $\mathbf{m}$ each time, with probability $p_{\mathbf{ m}}$, leading to an ensemble of quantum trajectories $\{p_{\mathbf m}, \rho_{\mathbf m} \}$. 

For a finite system of $N$ qubits, one expects that due to the continual extraction of information by the measurements, the system loses entropy over time, eventually becoming a pure state---a phenomenon known as {\it dynamical purification}~\cite{gullans_dynamical_2020}.
 Remarkably, the  time taken to  purify can differ drastically depending on the rate of measurements.
Heuristically, under strong/frequent local monitoring,   measurements are able to quickly read out all the information contained in the initial state, so that $\rho_{\mathbf{m}} \approx \ket{\psi_{\mathbf m}}\bra{\psi_{\mathbf m}}$ becomes pure quickly. As an extreme example, one may completely turn off the Hamiltonian or unitary evolution, and projectively measure all qubits simultaneously; then the system  purifies after just one timestep: given a bitstring of measurement outcomes $\mathbf{z}$, we have the collapse $\rho = \mathbb{I}/2^N \mapsto \rho_{\mathbf z} = \ketbra{\mathbf z}{\mathbf z}$, where $\{\ket{\mathbf z}\}$ is the computational basis.
On the other hand, for sufficiently weak/infrequent measurements, the dynamics manages to protect an extensive, or ``volume-law'', amount of information for a long time, captured by the von Neumann or  Reny\'i entropy:
$S(\rho_{\mathbf m})\propto N$.  
These two behaviors --- quick and slow dynamical purification, arising from the competition between information extraction by the measurements and information hiding by the dynamics
--- in fact give rise to genuine {\it phases} defined by differing information content present at late times, separated by a sharp transition at a critical measurement strength/rate.

More precisely, one may define a `purification time' $\tau_p$ as the minimum $t$ such that the trajectory-averaged entropy $\mathbb{E}_{\mathbf m}[ S(\rho_{\mathbf m}(t))]<\epsilon$, for some arbitrary small threshold $\epsilon >0$.
The purification phases can   be sharply distinguished in the large-system limit by the scaling of $\tau_p$ with $N$, namely $\tau_p\sim \log(N)$ in the ``pure phase'' (frequent measurements) and $\tau_p\sim \exp(N)$ in the ``mixed phase'' (infrequent measurements), see Fig.~\ref{fig:review}(d). 
In the mixed phase, the behavior of the trajectory-averaged entropy at late times can thus be captured well by the relation
\begin{equation}
    S(t) \sim e^{-2t/\tau_p}
    \label{eq:Sansatz}
\end{equation}
with $\tau_p\sim \exp(N)$, as was argued for in \cite{li_statistical_2021}.
This relation will turn out to play a key role in constraining deep thermalization times in this work. 
In Appendix~\ref{app:mixedphase}, we provide a review of the physics underlying such scaling behavior.

Finally, we highlight a formal connection between dynamical purification and the projected ensemble. 
In the discussion above, it is crucial to consider the average entropy of quantum trajectories $\mathbb{E}_{\mathbf m}[S(\rho_{\mathbf m})]$ as opposed to the entropy of the average of quantum trajectories $S(\mathbb{E}_{\mathbf m}[\rho_{\mathbf m}])$.
This is because purification phases are invisible to the first moment of the ensemble of trajectories, $\rho^{(1)} = \sum_{\mathbf m} p_{\mathbf m} \rho_{\mathbf m}$: this linear statistical mixture of the trajectories corresponds to a quantum channel acting on the input state $\rho(0) = \mathbb{I}/2^N$. Given that the quantum channel is unital\footnote{The identity operator $\mathbb{I}$ is separately invariant under unitary operations and under outcome-averaged projective measurements, so it is also invariant under their composition into a monitored circuit.}, the state $\rho(t)$ remains maximally mixed regardless of measurement rate. 
On the contrary, an order parameter that distinguishes the dynamical purification phases is e.g.~the trajectory-averaged purity, 
\begin{equation}
\sum_{\mathbf m} p_{\mathbf m} \Tr( \rho_{\mathbf m}^2 )
= \Tr( \rho^{(2)} \chi )\;,
\end{equation}
which in the r.h.s.~we write as the expectation of an observable ($\chi$, a swap between the two replicas of the system) on the second moment $\rho^{(2)}$ of the ensemble $\{p_{\mathbf m}, \rho_{\mathbf m} \}$ .
This is very similar to a ``higher-order'' observable Eq.~\eqref{eqn:Obs} probed by higher-moments of the projected ensemble, which go beyond observables captured in regular thermalization. 
As we will see in Sec.~\ref{sec:purification}, this connection between purification phases and deep thermalization can be made precise, and our understanding of monitored dynamics can be leveraged to obtain new results on the time taken for the emergence of state designs in the projected ensemble.


\section{Exact designs in dual-unitary circuits \label{sec:du} }

In this section, we present our first main result, that  random DU${}^+$ circuits (as defined in Sec.~\ref{sec:setup_std}), which lack dynamical purification physics in their space-time duals,  generically achieve an exact quantum state $k$-design for all $k$ at the thermalization time $t_1 = N_A$. This happens in the limit of infinitely large bath size $N_B \to \infty$.
We note of course that there are specific instances of DU${}^+$ circuits that do not exhibit this behavior, e.g., circuits composed of $\textsf{SWAP}$ gates, which are in $\mathfrak{DU}$ but do not generate entanglement, or the KIM at its dual-unitary but integrable point. However, what we show is that {\it almost all} DU${}^+$ circuits do so, greatly extending the scope of the results of Ref.~\cite{ho_exact_2022}, and showing that the phenomenology of emergent quantum state-designs in dynamics is generic, as far as this class of models goes.

\subsection{Setup and assumptions}

\begin{figure*}
\centering
\includegraphics[width=\textwidth]{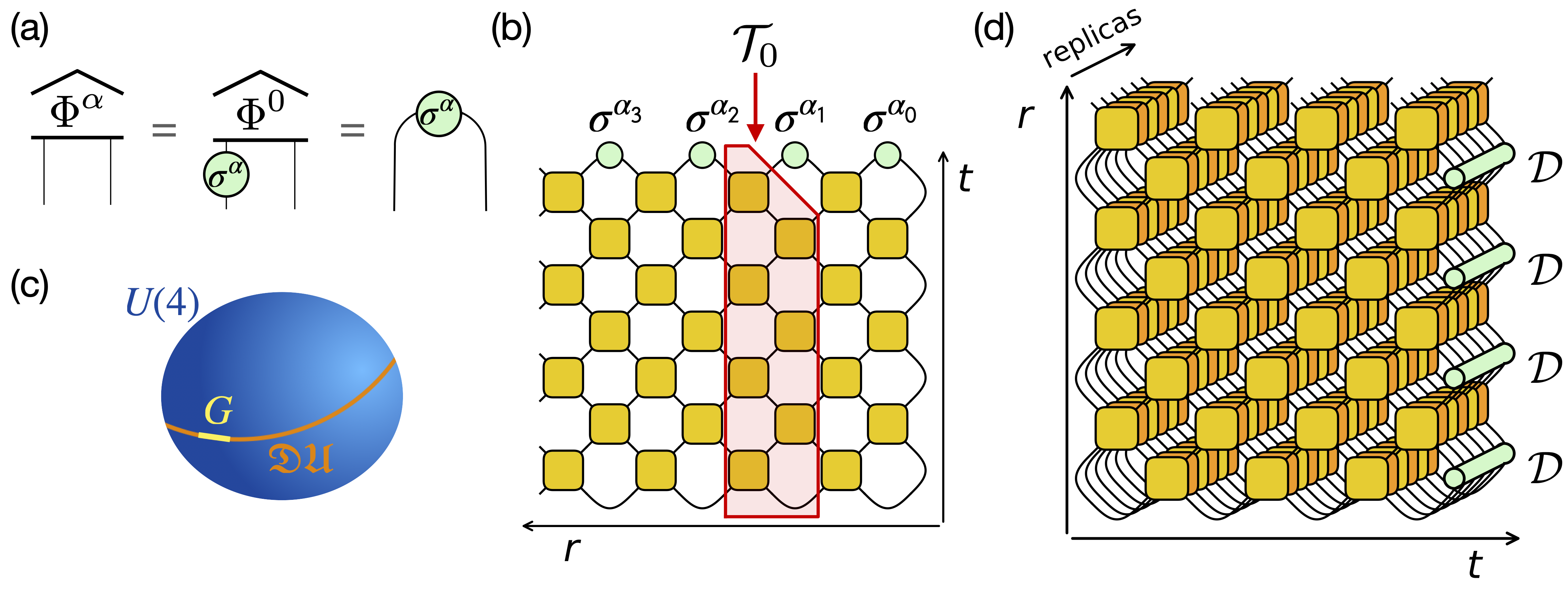}
\caption{Derivation of the projected ensemble in random DU${}^+$ circuits. 
(a) Projection onto the Bell state $\ket{\Phi^\alpha}$ is space-time dual to a single-qubit Pauli unitary $\sigma^\alpha$.
(b) Instance of a random dual-unitary circuit of depth $t=7$, truncated in the space direction at $N_B=8$. Squares represent dual-unitary gates, circles represent Pauli unitaries obtained from the Bell measurements. The shaded area represents a (unitary) transfer matrix $\mathcal{T}_0$.
(c) Schematic of $U(4)$ (the unitary group on two qubits), $\mathfrak{DU}$ (the submanifold of dual-unitary gates), and $G$ (an open subset of $\mathfrak{DU}$ where the gates making up each circuit instance may be sampled from). 
(d) Tensor network diagram for the state $\rho^{(k)}$ at $C$, for $k=3$, under space-time duality. The diagram is a stack of $k$ replicas of the unitary circuit (lighter gates) and $k$ replicas of its adjoint (darker gates). The average over measurement outcomes couples the replicas at an edge via a quantum channel $\mathcal{D}$ (green cylinders) repeatedly over dual-time $r$. \label{fig:du}}
\end{figure*}

Concretely, we consider brickwork circuits made of two-qubit gates in $\mathfrak{DU}$, with initial states and measurement bases chosen to ensure the unitarity of the transfer matrix $\mathcal{T}$. In particular, we fix an initial state composed of Bell pairs\footnote{This partition into pairs requires $N$ to be even; if $N$ is odd, we define $\ket{\psi_\text{init}} = \ket{\Phi^0}^{\otimes \lfloor N/2 \rfloor } \otimes \ket{0}$ instead.}, $\ket{\psi_\text{init}} = \ket{\Phi^0}^{\otimes N/2}$, with $\ket{\Phi^0} \equiv \frac{1}{\sqrt 2} (\ket{00} + \ket{11})$ and $N = N_A + N_B$ is the total number of qubits in the system, partitioned into subsystems $A$ and $B$. While we focus on this initial state for simplicity, any ``solvable'' matrix-product states (as defined in Ref.~\cite{piroli_exact_2020}) are amenable to the same treatment~\cite{claeys_emergent_2022}.
We also perform final measurements in the basis of two-qubit Bell states, i.e., pairs of neighboring qubits (that have {\it not} been coupled by the latest layer of unitary gates) are projectively measured in the orthonormal basis of Bell states $\ket{\Phi^\alpha} \equiv (\sigma^\alpha \otimes \mathbb{I}) \ket{\Phi^0}$. This choice of initial state and final measurement basis is in DU${}^+$, i.e., it guarantees that the transfer matrices $\mathcal{T}_\alpha$ (labeled by the possible outcomes $\alpha = 0, x, y, z$) are unitary. 
In fact, as shown diagrammatically in Fig.~\ref{fig:du}(a), such measurements dualize to single-qubit Pauli unitaries, giving
\begin{equation}
\mathcal{T}_\alpha =  \mathcal{T}_0 \sigma_t^\alpha,
\label{eq:t_alpha}
\end{equation}
where $\{ \sigma^\alpha_t\}$ are the Pauli matrices on site $t$ in the dual system. This is depicted schematically in Fig.~\ref{fig:du}(b). 
Note that we define $t$ as the circuit depth (i.e., the number of gates acting on each qubit), so that the size of $C$ is $|C| = t+1$; we number sites in $C$ as $\{0, 1, \dots t\}$. 
 
We will further consider circuits that are formed by sampling local gates $U$ independently and identically from $\mathfrak{DU}$ according to a probability distribution $P$.
We also require that the distribution $P$ has nonzero weight on an open subset $G\subset \mathfrak{DU}$, as sketched in Fig.~\ref{fig:du}(c). We impose no other requirements on $G$; in particular, it can be an arbitrarily small neighborhood of any $V\in \mathfrak{DU}$.
For future convenience, we introduce the a notion of ``circuit instance'' as follows.

\noindent {\bf Definition.} {\it A ``circuit instance'' (or just ``instance'') is a brickwork quantum circuit in 1+1 dimensions constructed by sampling two-qubit gates independently and identically from a probability distribution $P$ on $\mathfrak{DU}$. The quantum circuit has finite depth $t$ and is semi-infinite in space.}

Given an instance, we may truncate it at a finite length $N_B$, leaving dangling legs that define the timelike surface $C$, as in Fig.~\ref{fig:du}(b), and investigate the statistics of the quantum states produced at $C$ from the dual evolution as $N_B$ is increased.
As explained in Sec.~\ref{sec:setup_std}, it suffices to prove that these states are uniformly distributed, i.e.~form a state-design themselves, in order to show that the states comprising the projected ensemble on $A$ are also uniformly distributed (albeit on a different Hilbert space), provided that $|A| \leq |C|$. Below we focus on this situation, i.e. $N_A\leq t+1$.
For simplicity, we also assume that $N_A+t$ is odd in the following\footnote{This ensures that Bell measurements of $B$ can be performed without leaving an unpaired qubit near the the boundary between $A$ and $B$.}; the opposite case can be addressed with minor modifications.

\subsection{Spacetime-dual evolution}

We define the  ensemble of quantum states produced on $C$, with each state occuring with equal probability, as\footnote{Note that we have previously used this notation for the state produced at $A$; however, for the rest of this Section, we will focus uniquely on the state produced at $C$, so that there will be no ambiguity.} $\rho^{(k)}$.
By exchanging the roles of space and time, we can interpret the state $\rho^{(k)}$ as the output of a quantum channel acting on $k$ ``replicas'' of system $C$. 
This is shown diagrammatically in Fig.~\ref{fig:du}(d).
The channel is given by $\rho \mapsto \frac{1}{4} \sum_{\alpha} \mathcal{T}_\alpha^{\otimes k} \rho (\mathcal{T}_\alpha^{\otimes k})^\dagger$.
By using Eq.~\eqref{eq:t_alpha}, we can write the channel as $\rho \mapsto  \mathcal{U}^{\otimes k}\circ \mathcal{D} [\rho]$, 
where $\mathcal{U}$ is a unitary channel acting on a single replica as conjugation by $\mathcal{T}_0$, $\mathcal{U}[\rho] = \mathcal{T}_0 \rho \mathcal{T}_0^\dagger$, and $\mathcal{D}[\rho] = \frac{1}{4} \sum_\alpha (\sigma_t^\alpha)^{\otimes k} \rho (\sigma_t^\alpha)^{\otimes k}$ is a dissipative quantum channel coupling all replicas at one edge (qubit $t$), represented by the cylinders in Fig.~\ref{fig:du}(d). 
We will label the Kraus operators\footnote{Note that we choose to leave the prefactor $1/2$ outside the definition of $K_\alpha$ for future convenience. Therefore, the operators obey a modified normalization condition $\sum_\alpha K_\alpha^\dagger K_\alpha = 4\mathbb{I}$.} of this channel as $K_\alpha \equiv (\sigma_t^\alpha)^{\otimes k}$, $\alpha = 0,x,y,z$.

Henceforth we use $r$ to refer to ``time'' in this dual evolution: $r=1, \dots N_B$. Note that an iteration of the channel $\mathcal{U}^{\otimes k} \circ \mathcal{D}$ corresponds to $\Delta r = 2$.
Letting the initial state imposed by the open boundary condition in the original circuit be $\rho^{(k)}_0 \equiv (\ket{\psi_0}\bra{\psi_0})^{\otimes k}$, with $\ket{\psi_0} = \ket{\Phi^0}^{\otimes \lfloor \frac{t+1}{2} \rfloor}\otimes \ket{0}^{\otimes p}$ ($p\equiv (t+1) \mod 2$), the evolved state at dual-time $N_B$ is given by 
\begin{equation}
    \rho^{(k)}_{N_B} = \mathcal{U}^{\otimes k}_{N_B} \circ \mathcal{D} \circ \mathcal{U}^{\otimes k}_{N_B-2} \circ \mathcal{D} \circ \cdots \circ \mathcal{U}^{\otimes k}_2 \circ \mathcal{D} [\rho_0^{(k)}]\;,
    \label{eq:composition}
\end{equation}
where we have made explicit the fact that $\mathcal{U}$ depends on $r$ (via the sampling of i.i.d. random gates from $G\subset \mathfrak{DU}$).
Our goal now is to characterize the steady state of this sequence of channels: if it is the Haar moment $\rho_H^{(k)}$, we obtain an exact state $k$-design at $C$, and thus at $A$.

We may rewrite Eq.~\eqref{eq:composition} by evolving the Kraus operators in the Heisenberg picture, $K_\alpha(r) \equiv \mathcal{U}_{r\leftarrow 0}^{\otimes k} [K_\alpha]$, where $\mathcal{U}_{r\leftarrow 0} \equiv \mathcal{U}_r\circ \mathcal{U}_{r-2} \circ \cdots \circ \mathcal{U}_2$ is the unitary channel that implements the evolution between (dual) times 0 and $r$. 
Thus we have
\begin{align}
    \rho^{(k)}_{N_B} & = \mathcal{U}_{N_B \leftarrow 0}^{\otimes k} \circ \mathcal{D}_{N_B-2}^\prime \circ \mathcal{D}_{N_B-4}^\prime \circ \cdots \circ  \mathcal{D}_0^\prime [\rho_0^{(k)} ] 
    \label{eq:dprime_channels}\\
    \mathcal{D}^\prime_r[\rho] & =  \frac{1}{4} \sum_\alpha K_\alpha(r) \rho K_\alpha^\dagger(r)\;.
\end{align}
As the Haar moment $\rho_H^{(k)}$ is invariant under tensor-product unitaries, we may safely drop the unitary channel $\mathcal{U}_{N_B\leftarrow 0}^{\otimes k}$ and focus on the dissipative part, given by composition of the $r$-dependent $\mathcal{D}_r^\prime$ channels. 

\subsection{Convergence to the Haar moment}

In order to prove the emergence of exact state designs, we aim to show that (as $N_B \to\infty$) permutation operators are almost always the unique fixed points of this composition of channels~\cite{ho_exact_2022, claeys_emergent_2022}.
The outline of the proof is as follows: 
first, we show that every instance has a limit state $\rho_\infty^{(k)}$ (Appendix~\ref{app:limit_state});
then, we show $\rho_\infty^{(k)}$ must commute with the time-evolved Kraus operators of $\mathcal{D}$, $K_\alpha = (\sigma_{t}^\alpha)^{\otimes k}$; 
finally, we show that products of these time-evolved $K_\alpha$ operators generate the group $U(d_C)^{\otimes k} \equiv \{V^{\otimes k}:\ V\in U(d_C)\}$ {\it almost surely}, i.e., with probability that approaches 1 as $N_B\to\infty$. (Here $d_C = 2^{|C|} = 2^{t+1}$ is the Hilbert space dimension of $C$.)
From this, the formation of exact designs follows precisely from the same argument as in Ref.~\cite{ho_exact_2022}.

\noindent {\bf Theorem.} {\it The limit state $\rho^{(k)}_\infty$ is almost always the Haar moment $\rho^{(k)}_H$.}
\\

Before proceeding to the proof, let us stress that the above result  does not refer to the ensemble-averaged behavior of the random brickwork circuits; crucially, it holds at the level of a given circuit instance.
 
\noindent {\it Proof.} We bound the commutators between $\rho^{(k)}_\infty$ and the Kraus operators $\{K_\alpha(r)\}$ by triangle inequality, 
\begin{align}
\| [\rho^{(k)}_\infty, K_\alpha(r)] \|
& \leq \| [\rho^{(k)}_\infty - \rho^{(k)}_r, K_\alpha(r)] \| + \| [\rho^{(k)}_r, K_\alpha(r)] \| \nonumber \\
& \leq  2 \|\rho^{(k)}_\infty - \rho^{(k)}_r\|  + \| [\rho^{(k)}_r, K_\alpha(r)] \| \;,
\end{align}
where $\|\cdots \|$ is the trace norm and we used $\|AB\| \leq \|A\|_\infty \|B\|$ and $\|K_\alpha(r)\|_\infty = 1$.
By definition of limit state, there exists $r^\ast(\epsilon)$ such that for all $r>r^\ast(\epsilon)$ the first term is $\leq \epsilon$. 
The second term is also $\leq\epsilon$ due to a Lemma proven in Appendix~\ref{app:limit_state}.
Thus the limit state must commute, up to arbitrary accuracy $\epsilon$, with an infinite sequence of Heisenberg-evolved operators $\{K_\alpha(r):\ r> r^\ast(\epsilon)\}$. 

In Appendix~\ref{app:universal}, we show that, in almost all circuit instances, the set $\{ \mathcal{U}_{r\leftarrow 0}: r\in 2\mathbb{N}\}$ is dense in the space of unitary channels (the proof proceeds by showing any open set $G\subset\mathfrak{DU}$ is a universal gate set in the quantum computing sense~\cite{barenco_elementary_1995}).
Thus for any site $j\in\{0,\dots t\}$ and any two unit vectors $\mathbf{n}_{1,2}$ in $S^2$, there almost surely exist $r^\ast_{1,2}$ such that $\sigma_{t}^z(r^\ast_{1,2}) = \mathbf{n}_{1,2} \cdot \boldsymbol{\sigma}_j$ up to arbitrary accuracy $\epsilon>0$.
The limit state $\rho_\infty^{(k)}$ must therefore commute (almost always, up to accuracy $\epsilon$) with the product $K_z(r^\ast_1) K_z(r^\ast_2) = \exp(i\theta \mathbf{n}_3 \cdot \boldsymbol{\sigma}_j)^{\otimes k}$. The angle $\theta$ and direction $\mathbf{n}_3$ can be made arbitrary by choosing $\mathbf{n}_{1,2}$, thus generating the $k$-fold tensor powers of all single-qubit unitaries.
The same reasoning can be applied for entangling operations between any two qubits: for example there almost always exists $r^\ast$ such that $\sigma^{z}_{t}(r^\ast) \simeq e^{-i\theta \sigma^x_i \sigma^x_j} \sigma^z_i e^{i\theta \sigma^x_i \sigma^x_j}$, which is an entangling operation between qubits $i$ and $j$. 
Thus the limit state almost always commutes with a set of operators that generate the entire group $U(d_C)^{\otimes k}$. 
As the only operators that satisfy this condition exactly are permutations between the replicas (this follows from a mathematical result called the Schur-Weyl duality~\cite{marvian_generalization_2014, roberts_chaos_2017}), and the initial state $\rho_0^{(k)}$ is permutation-symmetric, we conclude that almost surely $\rho_\infty^{(k)} \propto \sum_{\pi \in S_k}\hat{\pi} $, where $\hat{\pi}$ is the operator on $\mathcal{H}_C^{\otimes k}$ that permutes replicas according to an element of the symmetric group $\pi \in S_k$; upon imposing unit trace, we conclude $\rho^{(k)} = \rho_H^{(k)}$ almost always. $\blacksquare$


We note that the result is independent of the gate set $G \subset \mathfrak{DU}$, provided this is an open subset. So we are free to take $G$ as a ball of radius $W$ around any dual-unitary gate $U\in \mathfrak{DU}$, take the limit $N_B\to \infty$ to recover the exact state designs at $C$ (and thus at $A$, following the discussion in Sec.~\ref{sec:setup_std}), and then make $W$ arbitrarily small, similar to the derivation of the spectral form factor in Ref.~\cite{bertini_random_2021}.
In other words, this result applies to arbitrary dual-unitary circuits with arbitrarily weak, spatiotemporally-uncorrelated disorder.


\section{Away from dual-unitarity: constraints from dynamical purification \label{sec:purification} } 

As soon as we break the DU${}^+$ conditions, either by perturbing the gates away from dual-unitarity or by modifying the initial state or final measurement basis, the above derivation fails. The transfer matrices become non-unitary, and the physics of monitored dynamics, reviewed in Sec.~\ref{sec:setup_monitored}, comes into play. In this Section we derive the consequences of this fact on the $k$-design times $t_k$.

\subsection{Bounding the purity of the $k$-th moment}

The connection to monitored dynamics is sharpened in the following inequality, which is one of the main results of this work.

\noindent {\bf Theorem.} {\it Consider a projected ensemble $\{(p(\mathbf{z}_1, \mathbf{z}_2), \ket{\psi_{\mathbf{z}_1, \mathbf{z}_2}})\}$ on a system of size $N$, where the projected states and probabilities are indexed by the bit-string $(\mathbf{z}_1, \mathbf{z}_2)$ such that $\mathbf{z}_1 \in\{0,1\}^r$ and $\mathbf{z}_2 \in\{0,1\}^{N-r}$ for some integer $r \geq 0$. Then, we have}
\begin{align}
    \Tr \left( {\rho^{(k)}}^2 \right) \geq \frac{1}{2^r} \left( \mathbb{E}_{\mathbf z_1} \Tr [(\rho_{\mathbf z_1}^{(1)})^2] \right)^k, \label{eq:main_theorem}
\end{align}
{\it where $\rho_{\mathbf z_1}^{(1)} = \sum_{\mathbf z_2} p(\mathbf z_2 | \mathbf z_1) \ket{\psi_{\mathbf z_1, \mathbf z_2}} \bra{\psi_{\mathbf z_1, \mathbf z_2}} $ is a density matrix defined in terms of the conditional probabilities $ p(\mathbf z_2 | \mathbf z_1) =  p(\mathbf z_2, \mathbf z_1)/ p(\mathbf z_1)$} and $\mathbb{E}_{\mathbf z_1}[\cdot] = \sum_{\mathbf z_1} p(\mathbf z_1) [\cdot ]$.

\noindent {\it Proof.}
We may write the l.h.s.~as
\begin{align}
\Tr \left( {\rho^{(k)}}^2 \right)
& = \sum_{\mathbf z_1, \mathbf z_1'} p(\mathbf z_1) p(\mathbf z_1') \sum_{\mathbf z_2, \mathbf z_2'} p(\mathbf z_2 | \mathbf z_1) p(\mathbf z_2' | \mathbf z_1') \nonumber \\
& \qquad \times \left| \bra{\psi_{\mathbf z_1, \mathbf z_2}} \psi_{\mathbf z_1', \mathbf z_2'}\rangle \right|^{2k} \nonumber \\
& =  \sum_{\mathbf z_1, \mathbf z_1'} p(\mathbf z_1) p(\mathbf z_1') \Tr(\rho_{\mathbf z_1}^{(k)} \rho_{\mathbf z_1'}^{(k)})
\end{align}
in terms of the $k$-th moments of the conditional ensembles. 
Then, by noting that all terms in the sum are non-negative and dropping off-diagonal terms, we have the inequality:
\begin{align}
\Tr \left( {\rho^{(k)}}^2 \right) 
& \geq \sum_{\mathbf z_1} p(\mathbf z_1)^2 \Tr[(\rho_{\mathbf z_1}^{(k)})^2] \;. \label{eq:deriv1}
\end{align}
At this point, we make use of two inequalties proven in Appendix~\ref{app:purity_bound}:
first, for any ensemble $\mathcal{E}$ we have the following bound between the purities of the first and $k$-th moment,
\begin{align}
\Tr({\rho^{(k)}}^2) 
& \geq \left[ \Tr({\rho^{(1)}}^2) \right]^k \;; \label{eq:purity_bound}
\end{align}
second, given a probability distribution $p(i)$ over $M$ elements and a non-negative function $f(i)$, we have
\begin{align}
\sum_i p(i)^2 f(i)^k
& \geq \frac{1}{M} \left( \sum_i p(i) f(i) \right)^k \;. \label{eq:holder}
\end{align}
Now, returning to Eq.~\eqref{eq:deriv1}, we have
\begin{align}
\Tr ({\rho^{(k)}}^2)
& \geq \sum_{\mathbf z_1} p(\mathbf z_1)^2 \left[ \Tr({\rho^{(1)}_{\mathbf z_1} }^2) \right]^k \nonumber \\
& \geq \frac{1}{2^r} \left( \sum_{\mathbf z_1} p(\mathbf z_1) \Tr({\rho^{(1)}_{\mathbf z_1} }^2) \right)^k \;,
\end{align}
where the first line follows from Eq.~\eqref{eq:purity_bound} and the second line follows from applying Eq.~\eqref{eq:holder} to the probability distribution $p(\mathbf z_1)$ (over $M=2^r$ elements) and the function $f(\mathbf z_1) = \Tr({\rho_{\mathbf z_1}^{(1)}}^2)$. $\blacksquare$

The result Eq.~\eqref{eq:main_theorem} places a constraint of the formation of state designs. Indeed, as we saw in Sec.~\ref{sec:proj_ens}, the l.h.s. equals the frame potential $F^{(k)}$ of the ensemble, which in turn can be used to formulate the $k$-design condition: one has $F^{(k)} \geq F_H^{(k)} \equiv \binom{2^{N_A}+k-1}{k}^{-1}$, with equality if and only if $\mathcal{E}$ forms a $k$-design.
If the ensemble $\mathcal{E}$ is to form a $k$-design, then the r.h.s. must not exceed the Haar frame potential $F_{H}^{(k)}$.

\begin{figure}
    \centering
    \includegraphics[width=\columnwidth]{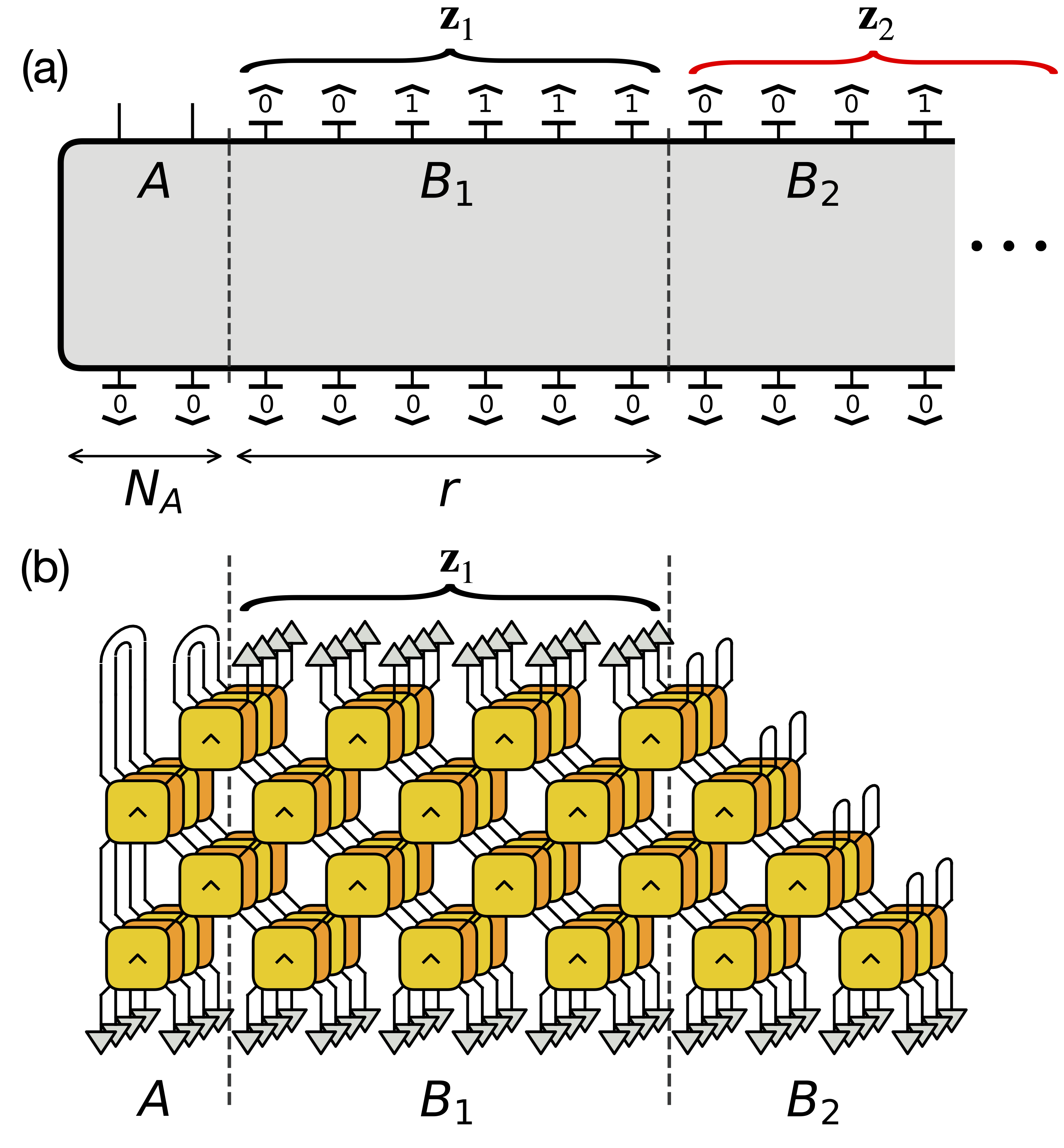}
    \caption{(a) Setup of Eq.~\eqref{eq:main_theorem}: the `bath' subsystem $B$ is partitioned into sub-regions $B_1$ (of size $r$) and $B_2$, yielding measurement outcomes $\mathbf{z}_1$ and $\mathbf{z}_2$ (example bitstrings are shown). From this setup one may form the conditional ensembles $\{ p(\mathbf{z}_2|\mathbf{z}_1), \ket{\psi_{\mathbf{z}_1,\mathbf{z}_2}}\}$ and their first moments $\rho^{(1)}_{\mathbf{z}_1}$. 
    (b) Tensor network diagram for the purity of $\rho_{\mathbf{z}_1}^{(1)}$ (up to a normalization factor $p(\mathbf{z}_1)^2$). Note the different contractions within subsystems $A$ and $B_2$. Due to unitarity, all but the nearest $t$ qubits in $B_2$ are elided. Viewed as an evolution in the space direction (from right to left), the diagram describes the dynamical purification of an initially mixed state.
    }
    \label{fig:purification}
\end{figure}

\subsection{Monitored dynamics and purification}

In the case of one-dimensional quench dynamics, the r.h.s. of Eq.~\eqref{eq:main_theorem} lends itself naturally to an interpretation in the language of monitored dynamics (reviewed in Sec.~\ref{sec:setup_monitored}). 
Let us consider a one-dimensional qubit chain partitioned into a finite subsystem $A$ (comprising the leftmost $N_A$ qubits) and its complement $B$, which serves as the bath to be measured in order to form the projected ensemble on $A$.
In addition, we split $B$ into two subsystems, $B_1$ comprising the $r$ qubits closest to $A$, and the remainder $B_2$, as illustrated in Fig.~\ref{fig:purification}(a).
We may then apply Eq.~\eqref{eq:main_theorem} to this situation to bound the purify of the $k$-th moment of the projected ensemble, with $\mathbf z_1\equiv \mathbf{z}_{B_1}$ and $\mathbf z_2 \equiv \mathbf{z}_{B_2}$. 

The term in the parenthesis of the r.h.s.~of Eq.~\eqref{eq:main_theorem} yields a  setup that has a direct interpretation in the language of monitored dynamics.
Specifically, the quantity 
\begin{equation}
\mathbb{E}_{\mathbf z_1} \Tr [(\rho_{\mathbf z_1}^{(1)})^2]
\end{equation}
is the purity of the density matrix $\rho^{(1)}_{\mathbf{z}_1}$, shown diagrammatically in Fig.~\ref{fig:purification}(b), ensemble-averaged over measurement outcomes $\mathbf{z}_{1}$.
What $\rho^{(1)}_{\mathbf{z}_1}$ in turn represents, upon invoking a space-time duality transformation, is an initially {\it highly-mixed} density matrix $\rho_\text{in}$ on a system of $|C|=t+1$ qubits (since the $B_2$ region is being traced out, as shown as the right boundary condition of Fig.~\ref{fig:purification}(b)) undergoing a quantum circuit evolution indexed by $\mathbf{z}_1 \in \{0,1\}^r$ in region $B_1$, before being mapped to $A$.
As reviewed in Sec.~\ref{sec:setup_monitored}, this quantum circuit can be segmented into transfer matrices, which are generically non-unitary and can be viewed as trajectories in a monitored evolution. 
In total, the depth of this monitored evolution is $r$. 
This is an instance of the problem of \emph{dynamical purification} of mixed states \cite{gullans_dynamical_2020}. 
Ref.~\cite{ippoliti_fractal_2022} showed that space-time duals of unitary circuits are generically in the \emph{mixed phase} of the dynamical purification problem, where for a system of $N$ qubits one has $\tau_p \sim \exp(N)$~\cite{gullans_dynamical_2020, fidkowski_how_2021, li_statistical_2021}. More directly, it is possible to show (see Appendix~\ref{app:mixedphase}) that the case at hand is exactly equivalent to a unitary circuit on $t+1$ qubits with 2 measurements per time step---as $t$ grows, the density of measurements vanishes as $\sim 2/t$, bringing the dynamics squarely in the mixed phase.

Thus, by interpreting the r.h.s.~of Eq.~\eqref{eq:main_theorem} as arising from spacetime-dual purification dynamics, we can write it as $2^{-r -k S_2^{(a)}(r) }$, where $S_2^{(a)}$ is the ``annealed average'' of the second R\'enyi entropies (measured in bits) of the ensemble of ``trajectories'' labeled by measurement outcomes $\mathbf{z}_1$, i.e., $S_2^{(a)} = -\log_2 \mathbb{E}_{\mathbf{z}_1} \Tr[(\rho_{\mathbf z_1}^{(1)})^2]$. 
In the following we simply denote this quantity by $S$
and assume the scaling relation in the mixed phase, Eq.~\eqref{eq:Sansatz}, argued for in Sec.~\ref{sec:setup_monitored} and the Appendix~\ref{app:mixedphase}, 
upon invoking the dictionary $N \leftrightarrow t$, $t \leftrightarrow r$ and $\tau_p \sim \exp(N) \leftrightarrow \xi_p \sim \exp(t)$ to relate variables of the standard setup of dynamical purification in Sec.~\ref{sec:setup_monitored} to those in our present problem respectively. 
Namely, these are the system `size', circuit `depth', and purification scales (in the mixed phase).
Concretely, the ansatz we use for $S$ at large $r$ will thus be
\begin{align}
    S(r) \sim e^{-2r/\xi_p},
\end{align}
where $\xi_p$ is a ``purification length scale'' that diverges exponentially in $t$.
It will be helpful to define the \emph{purification velocity} 
\begin{equation}
v_p \equiv \lim_{t\to\infty} \frac{1}{t} \log_2 \xi_p(t);
\end{equation}
we note this quantity can be interpreted as the line tension of a domain wall in a statistical mechanical description of purification phases which is discussed in Appendix~\ref{app:mixedphase}.  

\subsection{Constraints on the design times}

To recapitulate, Eq.~\eqref{eq:main_theorem} relates the formation of state designs (represented by the frame potential $F^{(k)}$ in the l.h.s.) to the problem of dynamical purification in the spacetime-dual dynamics (represented by the $r$-dependent average purity in the r.h.s.). 
This allows us to derive constraints on the formation of state designs by using insights about dynamical purification. Heuristically, the finite memory time of monitored dynamics may cause measurement outcomes very far away from $A$ to be ``forgotten'' and thus effectively limit the size of the projected ensemble, which obstructs the formation of high designs. This is in contrast to DU${}^+$ circuits, where the dual evolution is unitary and thus has perfect memory -- measurement outcomes arbitrarily far away always have an  effect on the state in $A$. 

Concretely, we may rewrite Eq.~\eqref{eq:main_theorem} as
\begin{equation}
\ln(1/F^{(k)}) \leq \ln(2)[r+ kS(r)] \;.
\label{eq:logbound}
\end{equation}
This is a family of bounds parametrized by $r\in \mathbb{N}$, all of which must be satisfied. We can replace this family of bounds by the most stringent one, i.e., minimize the r.h.s.~over $r$. This yields $r = r^\ast(k)$, where $r^\ast$ solves\footnote{We make $r$ continuous and interpolate $S(r)$ to a smooth function of a real variable.} $S'(r^\ast) = -1/k$. 
Here we invoke our ansatz motivated by dynamical purification: we assume $S(r)$ monotonically decreases and asymptotes to 0. This implies that its derivative $S'(r)<0$ also asymptotes to 0, so $r^\ast(k)$ diverges as $k\to\infty$. This ensures that, if we take $k$ large enough, $r^\ast$ lies in the domain of applicability of the large-$r$ ansatz $S(r) \sim e^{-2r/\xi_p}$. 
Minimizing the r.h.s.~under this ansatz gives $r^\ast = \frac{\xi_p}{2} \ln \frac{2k}{\xi_p }$, and thus the bound
\begin{equation}
\ln (1/F^{(k)}) \leq \xi_p \frac{\ln(2)}{2} \ln \left( \frac{2ek}{\xi_p} \right) \;.
\end{equation}

The $\epsilon$-approximate $k$-design condition $\Delta_2^{(k)}\leq \epsilon$ (see Eq.~\eqref{eq:approx_design_time}) yields $F^{(k)}\leq (1+\epsilon^2)F_H^{(k)}$, where $F_H^{(k)} \equiv \binom{2^{N_A} + k-1}{k}^{-1}$ is the frame potential of the Haar ensemble, as in Eq.~\eqref{eq:haar_fp}. Thus
\begin{equation}
\ln \binom{2^{N_A}+k-1}{k} - \ln(1+\epsilon^2) \leq \xi_p \frac{\ln(2)}{2} \ln \left( \frac{2ek}{\xi_p} \right) \;. \label{eq:bound_finite_k}
\end{equation}
We may now take $k\to\infty$ (which is consistent with the regime of applicability of our ansatz for $S(r)$). In this limit, using $\ln \binom{k+n}{k} = n\ln(k) -\ln(n!) + O(1/k)$ we have that both sides of Eq.~\eqref{eq:bound_finite_k} diverge as $\ln(k)$; thus we obtain a bound on the prefactors of $\ln(k)$,
\begin{equation}
2^{N_A}-1 \leq \xi_p(t) \frac{\ln(2)}{2} \;. \label{eq:bound_xi_NA}
\end{equation}
This bound holds for any value of $N_A$, and constrains the time needed to form infinitely high designs, $t_\infty \equiv \lim_{k \to\infty} t_k$.
Finally, for large $t$ and $N_A$, Eq.~\eqref{eq:bound_xi_NA} reduces to 
\begin{equation}
t_\infty \geq \frac{N_A}{v_p} \;.
\end{equation}
In the same limit (large $N_A$), the thermalization time $t_1$ defines the ``entanglement velocity'', $v_E \equiv {N_A} / {t_1}$. Using this definition to eliminate $N_A$ yields the inequality
\begin{equation}
t_\infty \geq \frac{v_E}{v_p} t_1 \;,
\label{eq:separation}
\end{equation}
which is another main result of our work.

Eq.~\eqref{eq:separation} shows that, whenever $v_p < v_E$, we have a guaranteed minimum separation between $t_1$ (governing the formation of a 1-design, or regular thermalization) and $t_\infty$ (governing the formation of high designs, or deep thermalization).

\subsection{Tuning the purification velocity \label{sec:tuning_vp} }

As $t_\infty\geq t_1$ is always true, in order for Eq.~\eqref{eq:separation} to be nontrivial, one must be able to tune $v_p$ below $v_E$.
Here we discuss some examples in which this can be achieved, displaying a genuine separation between design times.

We consider DU circuits with solvable initial states, where it is known that $v_E = 1$~\cite{piroli_exact_2020}. 
By choosing the gates and the final measurement basis, it is possible to vary $v_p$ significantly while $v_E$ is pinned to 1. 
Evidence of this can be seen by simulating random DU circuits with variable gate sets and measurement bases. In particular, we consider measurement bases that interpolate smoothly between the Bell basis and the computational basis, parametrized by $\mu \in [0,1]$.
Specifically, we project pairs of qubits onto pure states $\ket{\psi_\mu^a} \propto (I\pm \mu\sigma^z_1) \ket{\Phi^a}$, where $\{\ket{\Phi^a}:a=0,x,y,z\}$ is the Bell basis. Note that for $\mu=0$ we recover exactly the Bell basis, while for $\mu=1$ we obtain a disentangled basis $\{\ket{00}, \ket{01},\ket{10},\ket{11}\}$; the entanglement of the basis states decreases monotonically with $\mu$.
 
\begin{figure}
    \centering
    \includegraphics[width=\columnwidth]{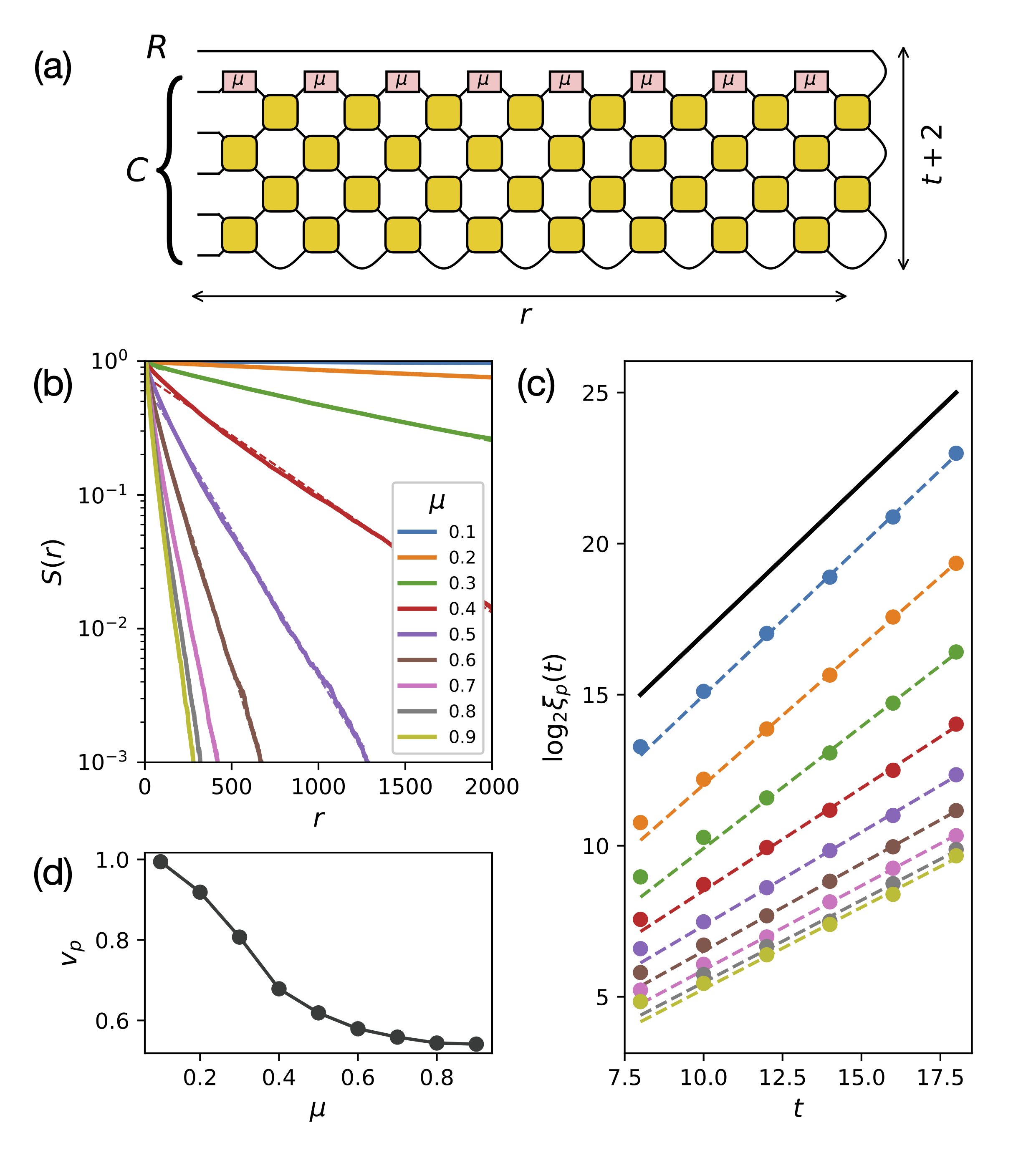}
    \caption{Tuning the purification velocity $v_p$ in DU circuits. 
    (a) Circuit setup for the numerical simulations: an initial state is prepared on $t+2$ qubits (right end), with one qubit set aside as a reference $R$; the remaining $t+1$ qubits, making up system $C$, evolve under DU gates (yellows squares) and measurements (boxes labelled by $\mu$) at an edge qubit. This example shows $t=4$. The entanglement between $R$ and $C$ is computed as a function of dual-time $r$.
    (b) Entropy of the reference qubit $S(r)$ for a $t=12$ and varying measurement scheme $\mu$ ($\mu=0$ gives Bell measurements, $\mu=1$ gives $Z$ measurements). We use the annealed average of the second Renyi entropy, $S\equiv S_2^{(a)}$, as in Eq.~\eqref{eq:main_theorem}. Dashed lines are exponential fits, $S(r)\sim e^{-2r/\xi_p(t)}$.
    Numerical data is obtained from exact simulations and averaged over between 500 and 10000 realizations, depending on system size.
    (c) Purification lengthscale $\xi_p(t)$ vs $t$, extracted from fits as shown in (b). The values of $\mu$ are the same and indicated by the same colors. The solid line has unit slope, for reference. The dashed lines are exponential fits $\xi(t)\sim 2^{v_p t}$.
    (d) Extracted values of $v_p$ vs $\mu$. We observe $v_p\to 1$ as $\mu\to0$. 
    }
    \label{fig:tuning_vp}
\end{figure}

We simulate the spacetime-dual (monitored) dynamics of circuits formed from random DU gates (specifically, the single-qubit gates $r,s,u,v$ are Haar-random in $SU(2)$ and $J$ is uniformly distributed in $[0,\pi/4]$), acting on $t+2$ qubits, one of which serves as a reference $R$ (i.e., is initially entangled with the rest and is not touched afterwards). As shown in Fig.~\ref{fig:tuning_vp}(a), this reproduces the setup of Eq.~\eqref{eq:main_theorem} with a minimal subsystem $B_2$ (the single reference qubit $R$), a subsystem $B_1$ of length $r$ (duration of the dual time evolution), and a subsystem $C$ of $t+1$ qubits.
As a result of the measurements on $B_1$, the evolution is generally monitored (other than at $\mu = 0$, where we measure $B_1$ in the Bell basis and thus fulfill the DU${}^+$ conditions). 
The family of measurement bases we introduced above, parametrized by $\mu\in[0,1]$, dualizes to {\it weak measurements} on the last system qubit; these are given by Kraus operators $K_\pm \propto I \pm \mu \sigma^z$ followed by single-qubit Pauli gates. Note that for $\mu = 1$ we recover strong $\sigma^z$ measurements, while for $\mu = 0$ we obtain purely unitary operations.

We compute the entropy $S(r)$ of the reference qubit as a function of $r$, and observe an exponential decay $S(r)\sim e^{-2r/\xi_p(t)}$ (Fig.~\ref{fig:tuning_vp}(b)). The purification length scale $\xi_p(t)$ is found to be consistent with the expected behavior $\sim \exp(t)$ in the numerically explored range, $8\leq t\leq 18$ (Fig.~\ref{fig:tuning_vp}(c)).
Moreover, as the ``measurement strength'' $\mu$ is tuned we find a significant variation of the purification velocity $v_p$, extracted from exponential fits to $\xi_p(t)$. Fig.~\ref{fig:tuning_vp}(d) shows that $v_p$ approaches $1$ from below as $\mu \to 0$. This is in line with the results of Sec.~\ref{sec:du}: $\mu=0$ corresponds to the DU${}^+$ regime where we have proven the instantaneous emergence of all state designs, i.e.~$t_k = t_1 = N_A/v_E$ for all $k$; this immediately implies that the bound in Eq.~\eqref{eq:separation} must be trivial, i.e.~$v_p \geq v_E$. Recalling that in this case $v_E=1$, we must have $v_p \geq 1$ as $\mu\to 0$.
For $\mu \neq 0$, we see $v_p < 1$, indicating a separation between the regular thermalization time and higher state-design formation time.

\begin{figure}
    \centering
    \includegraphics[width=\columnwidth]{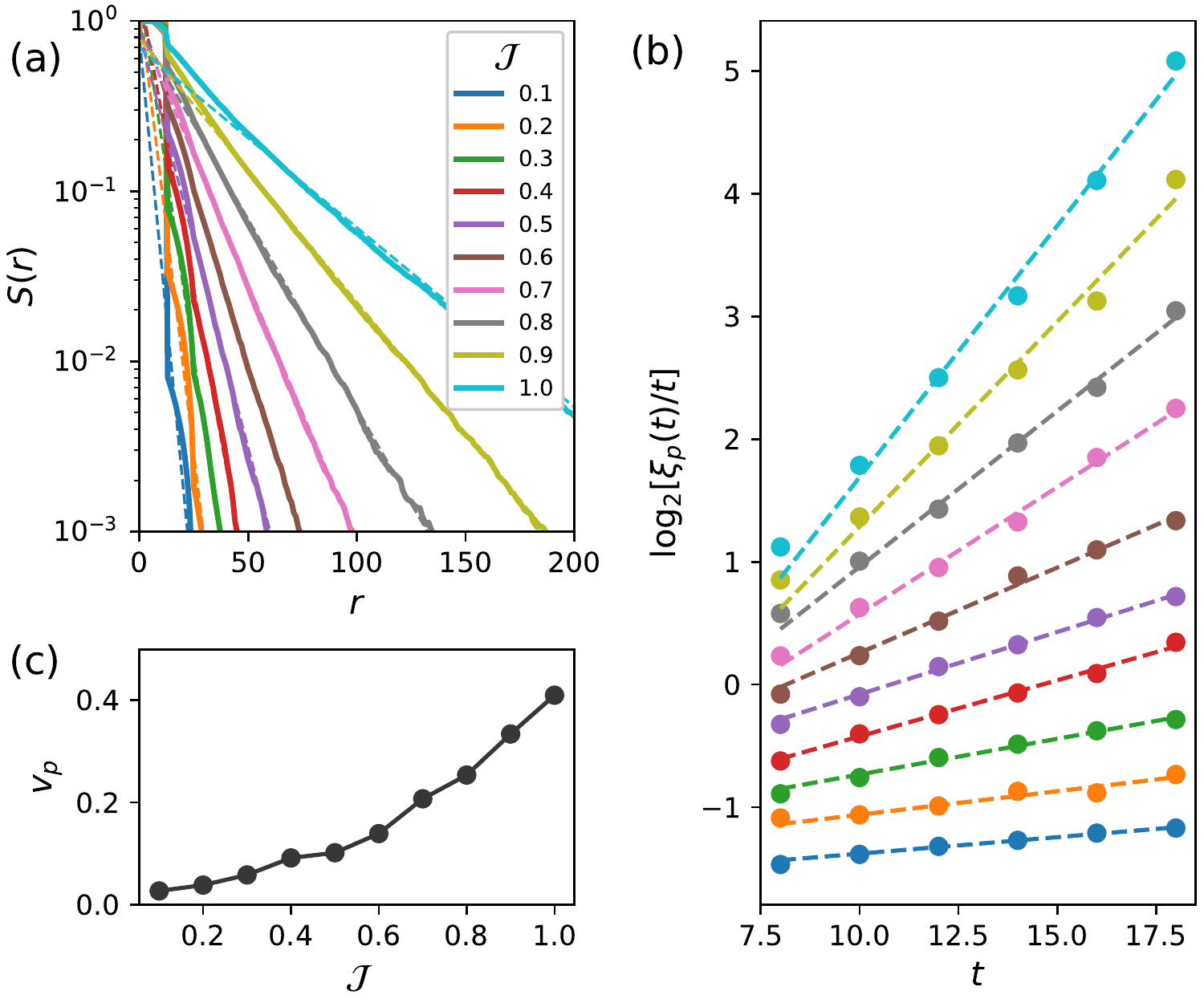}
    \caption{(a) Same setup as Fig.~\ref{fig:tuning_vp}, but for fixed measurement basis (computational basis, corresponding to $\mu=1$) and variable interaction strength $\mathcal{J}$.     Numerical data is obtained from exact simulations and averaged over between 200 and 10000 realizations, depending on system size.
    (b) Purification length scale $\xi_p(t)$, extracted from fits as shown in (a). The values of $\mathcal{J}$ are the same and indicated by the same colors. We plot the normalized quantity $\xi_p(t)/t$ to eliminate a prefactor of $t$ that arises in the limit of $\mathcal{J}\to 0$.
    The dashed lines are exponential fits $\xi_p(t)/t \sim 2^{v_p t}$.
    (c) Extracted values of $v_p$ vs $\mathcal{J}$. We observe $v_p\to 0$ as $\mathcal{J}\to0$. 
    }
    \label{fig:tuning_vp2}
\end{figure}

In fact, the separation between $v_p$ and $v_E$, and thus between $t_1$ and $t_\infty$, can be made arbitrarily large. To this end, we consider random DU circuits with gates sampled from a gate set 
$G = \{ r_1s_2 {\sf SWAP} e^{-iJ Z_1 Z_2} u_1v_2: |J|<\frac{\pi}{4} \mathcal{J} \} \subset \mathfrak{DU}$, where $r,s,u,v\in SU(2)$ are unconstrained and the parameter $\mathcal J$ plays the role of an interaction strength.
Namely, as $\mathcal{J}\to 0$ we recover non-interacting {\sf SWAP} circuits, while for $\mathcal{J}=1$ we have $G = \mathfrak{DU}$. 
We numerically simulate the same setup considered in Fig.~\ref{fig:tuning_vp}, although with fixed $\mu=1$ (i.e., measurements in the computational basis) and variable $\mathcal{J}$. Results, in Fig.~\ref{fig:tuning_vp2}, show that $v_p$ approaches 0 as $\mathcal{J}\to 0$. Note that this limit yields a ``critical'' purification, $\xi_p\propto t$, owing to the fact that measurements happen only at an edge of the system and qubits must travel (ballistically) to that edge in order to be measured and purified. While this algebraic prefactor does not modify the definition of $v_p$, it does impact numerical fits on a limited dynamical range when $v_p$ is small; for this reason, we extract $v_p$ from fits to the normalized quantity $\xi_p/t$, shown in Fig.~\ref{fig:tuning_vp2}(b).

These numerical results ($0\leq v_p \leq 1$, with $v_p=1$ at DU${}^+$ and $v_p=0$ in non-interacting circuits) are in line with expectations from the theory of measurement-induced entanglement transitions, reviewed in Sec.~\ref{sec:setup_monitored}. In particular, they are suggestive of the fact that $v_p$ should correspond to a \emph{line tension} in a statistical-mechanical description of entanglement in the random circuit\cite{jian_measurement-induced_2020, choi_quantum_2020, li_statistical_2021, li_entanglement_2021}. Universal scaling arguments therein give reason to believe that the same line tension controls both $v_p$ and the polynomial-depth ``plateau'' of the entropy density (i.e. the order parameter of the mixed phase\cite{gullans_dynamical_2020}) during dynamical purification. 
Since the latter is between 0 and $1$ (if we measure the entropy in bits), this correspondence would automatically yield the observed range of $v_p$. Finally, as one takes the measurement strength to zero one expects the maximum of the entropy density to be achieved, and thus the equality $v_p = 1$; on the contrary, taking the interaction strength to zero achieves the minimum value $v_p=0$, corresponding to an instability of the mixed phase towards critical purification~\cite{ippoliti_postselection-free_2021}.


\section{Discussion \label{sec:discussion}}

We have analyzed the physics of formation of quantum state designs in the projected ensemble, a novel emergent random matrix universality, arising  from states generated under the dynamics of one-dimensional quantum systems, that we modeled by unitary circuits. The formation of state $k$-designs provides a hierachy of time scales, $\{t_k: k\in \mathbb{N} \}$, that include the thermalization time $t_1$, but also other scales $t_{k>1}$ whose physical significance is still largely unclear. 
By connecting this problem to the phenomenology of monitored systems, through a space-time duality approach, we have highlighted the important role of dynamical purification in differentiating the formation of high designs from regular thermalization.

First, we have shown that in the absence of dynamical purification (achieved in a class of models that we named DU${}^+$, characterized by dual-unitary gates and compatible initial states and final measurements), one generically obtains {\it exact} state designs, and all time scales $t_k$ collapse onto the thermalization time $t_1$. This extends the phenomenology recently discovered in a Floquet Ising model~\cite{ho_exact_2022} to a wide class of spatiotemporally-disordered models.
Secondly, we have shown that, upon breaking the DU${}^+$ conditions and restoring dynamical purification, it is possible to derive nontrivial constraints on the design times, in particular a separation between the thermalization time $t_1$ and the time needed to form ultra-high designs, $t_\infty = \lim_{k\to\infty} t_k$.

The bound we have derived, $t_\infty > (v_E/v_p) t_1$, is nontrivial whenever the ``purification velocity'' $v_p$ of the circuit is smaller than its ``entanglement velocity'' $v_E$. We have shown, with physical arguments corroborated by numerical simulations, that this can be realized in a wide class of models, and that the separation can in fact be made arbitrarily large---e.g. one can take $v_p\to 0$ with constant $v_E=1$, as shown in Fig.~\ref{fig:tuning_vp2}.
Nonetheless, our bound for $t_k$ is only $O(1)$ in $k\to\infty$. It is an interesting open question whether our bounds are tight; in particular, is it possible to derive (either from dynamical purification arguments or from independent routes) bounds that {\it diverge} in $k$?
On a discrete set of $d$ elements, a probability distribution only has a finite number of independent moments (namely $d-1$), therefore $k$-designs beyond $k=d$ are all trivially formed at the same time, and the limit $k\to\infty$ cannot yield divergent time scales. 
A $d$-dimensional Hilbert space behaves somewhat similarly, in the sense that $k$-designs beyond $k=d$ are ``close'' to each other in a specific sense~\cite{roberts_chaos_2017}.
However it is not clear {\it a priori} whether a divergent bound, $t_k>f(k)$ for some $f$ obeying $\lim_{k\to\infty}f(k)=\infty$, can be ruled out as a consequence of this.

As we have shown, the formation of higher state-designs in the projected ensemble probes physics that go beyond regular quantum thermalization. We conclude with a set of exciting directions for future research, which aims to better understand the connection between this novel phenomenon and various other aspects of non-equilibrium dynamics:

{\it Scrambling.} The models displaying the strongest separation of time scales between $t_1$ and $t_\infty$ are notably weakly-interacting, suggesting a connection to quantum information scrambling\cite{hayden_black_2007, shenker_black_2014, hosur_chaos_2016, roberts_chaos_2017, von_keyserlingk_operator_2018, nahum_operator_2018, swingle_measuring_2016, mi_information_2021}.
It is thus interesting to sharpen the connection between the formation of high designs and the well-established diagnostics of scrambling, such as out-of-time-ordered correlators~\cite{von_keyserlingk_operator_2018, nahum_operator_2018, swingle_measuring_2016} and the tripartite mutual information of the circuit~\cite{hosur_chaos_2016, sunderhauf_quantum_2019}. In particular, what is the relationship between the ``butterfly velocity'' $v_B$ that controls scrambling and the ``purification velocity'' $v_p$ that governs high designs? While the former is a property of the bulk circuit and of the initial state, the latter is also dependent on a choice of final measurement basis on the ``bath''; thus presumably any relationship between the two ought to involve an average or optimization of $v_p$ over the choice of local measurement basis.

{\it Chaos/Ergodicity.} There are striking similarities between the derivation of the exact RMT spectral form factor (in the KIM~\cite{bertini_exact_2018} and subsequently in generic DU circuits~\cite{bertini_random_2021}) and the derivation of exact state designs (also in the KIM~\cite{ho_exact_2022} and extended to generic DU${}^+$ circuits in this work), both of which leverage a space-time duality mapping in order to derive different incarnations of RMT behavior in microscopic models of dynamics. There are also crucial distinctions, however, such as the need for a Floquet evolution in the former and the dependence on initial and final states in the latter. 
Another important distinction is that in the former, an ensemble average (e.g., over disorder realizations) is required because the spectral form factor is not self-averaging~\cite{gharibyan_onset_2018}, while in the latter, the result appears already at the level of a {\it single} quantum many-body state.
It is intriguing to speculate on a deeper connection between the two phenomena, such as what role quantum chaos might play in the formation of high state designs in the projected ensemble, or even whether the emergence of quantum state designs can be used as an alternative definition of many-body quantum  chaos.

{\it Integrability.} Models with $v_E>0$ and $v_p=0$ thermalize, but fail to form high designs in the projected ensemble. An example of this is a {\sf SWAP} circuit acting on an initial state of Bell pairs: this model generates entanglement between subsystems by transporting entangled excitations. Interestingly, this can be viewed as a cartoon model of post-quench entanglement generation in integrable models~\cite{alba_entanglement_2017, calabrese_entanglement_2020}. Does this behavior (thermalization without higher design fomation) generalize to non-trivial, interacting models? Exploring this question in tractable circuit models of interacting-integrable systems~\cite{gopalakrishnan_operator_2018, klobas_exact_2021, buca_rule_2021} is an interesting next step.

{\it Teleportation/complexity transition.}
Lastly, we note that our results apply only to models in one spatial dimension, where the mutual information between two spatially-separated degrees of freedom (upon measuring all the others) decays exponentially with distance. This is at the core of our ansatz for the scaling of entropy $S(r)$ that enables the derivation of our bound between $t_1$ and $t_\infty$, and is a consequence of the absence of long-range order in one dimension. However, in higher dimension one generically expects a finite-depth transition where said mutual information becomes finite even at infinite distance---a phenomenon that can be interpreted as a {\it teleportation transition}~\cite{bao_finite_2022}, closely related to a phase transition in the complexity of sampling the output of shallow two-dimensional circuits~\cite{napp_efficient_2022}.
An exciting question is how this transition might impact the projected ensemble on states formed from quench dynamics in higher-dimensional systems. For example, can the teleportation/complexity transition be detected by analyzing quantitative features of the formation of high designs? In the ``ordered'' phase, where our arguments based on dynamical purification cease to apply, is there still generically a separation between the time scales $t_1$ and $t_\infty$? 
Recent work on a random-matrix model of deep thermalization~\cite{ippoliti_solvable_2022} found a finite separation $t_\infty / t_1 \simeq 2$, suggesting that the DU${}^+$ result ($t_1 = t_\infty$) ought to be non-generic even in higher dimension.
However, it is interesting to speculate that generic locally-interacting systems in dimension 2 and higher might match the random-matrix result, in this sense reaching deep thermalization as fast as possible after regular thermalization.

\acknowledgments

We thank S.~Choi, T.~Rakovszky and V.~Khemani for discussions and for previous collaborations on related topics. M.~I.~thanks N.~Hunter-Jones for helpful discussions on quantum state designs.
We are especially grateful to D. Mark for pointing out to us the proof of monotonicity of design times in Appendix~\ref{app:monotonicity}.
M.~I.~is supported by the Gordon and Betty Moore Foundation's grant GBMF8686 and by the Defense Advanced Research Projects Agency (DARPA) via the DRINQS program. The views, opinions and/or findings expressed are those of the authors and should not be interpreted as representing the official views or policies of the Department of Defense or the U.S. Government.
W.~W.~H.~is supported in part by the Stanford Institute of Theoretical Physics and in part by the National University of Singapore  start-up grants A-8000599-00-00 and A-8000599-01-00.
Numerical simulations were performed on Stanford Research Computing Center's Sherlock cluster.
This project originated from discussions at the KITP programs {\it ``Energy and Information Transport in Non-Equilibrium Quantum Systems''} and {\it ``Non-Equilibrium Universality: From Classical to Quantum and Back''}; KITP is supported by the National Science Foundation under Grant No.~NSF PHY-1748958.

\onecolumngrid
\appendix

\section{Monotonicity of distances and design times \label{app:monotonicity}}

In this Appendix, we show that the distances $\Delta^{(k)}_\alpha$ of the $k$-th moment of the projected ensemble to the uniform ensemble,
\begin{align}
    \Delta^{(k)}_\alpha \equiv \frac{\| \rho^{(k)} - \rho^{(k)}_H \|_\alpha}{ \|\rho^{(k)}_H \|_\alpha},
\end{align}
obey monotonicity for Schatten indices $\alpha \geq 1$: 
\begin{align}
    \Delta^{(k+1)}_\alpha \geq \Delta^{(k)}_\alpha.
\end{align}

A consequence of monotonicity is that design times $\{ t_{k,\alpha}\}$ as defined in Eq.~\eqref{eq:approx_design_time} also obey monotonicity, i.e., they are non-decreasing in $k$. This is a desirable property that allows for the interpretation of the design times as a sequence of timescales describing progressively ``deeper'' levels of thermalization.

{\bf Proof.}
Since $\rho^{(k)}$ is a Hermitian operator associated with a density matrix, it has a spectral decomposition 
\begin{align}
    \rho^{(k)} = \sum_{i=1}^{D_k} \lambda_i^{(k)} |\Lambda_i^{(k)}\rangle \langle \Lambda_i^{(k)} |,
\end{align}
where eigenvalues $\lambda_i^{(k)} \geq 0$ and  eigenvectors $\{  |\Lambda_i^{(k)}\rangle \}_{i=1}^{D_k}$ form an orthonormal basis for the symmetric subspace of $\mathcal{H}_A^{\otimes k}$ with dimension $D_k = \binom{d_A + k - 1 }{k}$. Here $d_A$ is the dimension of $\mathcal{H}_A$. Therefore, 
\begin{align}
    \rho^{(k)}_H = \sum_{i=1}^{D_k} \frac{1}{D_k} |\Lambda_i^{(k)}\rangle \langle \Lambda_i^{(k)}|.
\end{align}
Define
\begin{align}
    h_{ij} \equiv \langle \Lambda_i^{(k)}| \Tr_{k+1} \left[ |\Lambda_j^{(k+1)}\rangle \langle \Lambda_j^{(k+1)} | \right] |\Lambda_i^{(k)}\rangle, 
\end{align}
where the trace is over the $(k+1)$-th copy of the Hilbert space, and 
where $1 \leq i \leq D_k$ and $1 \leq j \leq D_{k+1}$. One can immediately verify that $h_{ij} \geq 0$. Furthermore, 
\begin{align}
    \sum_{i=1}^{D_k} h_{ij} & = \Tr \left[ \Pi_\text{symm}^{(k)} \otimes \mathbb{I}_{k+1} |\Lambda_j^{(k+1)}\rangle \langle \Lambda_j^{(k+1)} | \right] = 1, \\
    \sum_{j=1}^{D_{k+1}} h_{ij} & = \langle \Lambda_i^{(k)} | \Tr_{k+1} \Pi^{(k+1)}_\text{symm} | \Lambda_i^{(k)}\rangle 
    = \langle \Lambda_i^{(k)}  | \left(\frac{D_{k+1}}{D_k} \right) \Pi^{(k)}_\text{symm} | \Lambda_i^{(k)}\rangle 
    = \left(\frac{D_{k+1}}{D_k} \right).
\end{align}
We have
\begin{align}
    \| \rho^{(k)} - \rho^{(k)}_H \|_\alpha^\alpha & = \sum_{i=1}^{D_k} \left| \lambda_i^{(k)} - \frac{1}{D_k} \right|^\alpha \nonumber \\
    & = \sum_{i=1}^{D_k} \left|  \langle \Lambda_i^{(k)} | \left( \rho^{(k)} - \rho^{(k)}_H \right)|\Lambda_i^{(k)} \rangle \right|^\alpha \nonumber \\
    & = \sum_{i=1}^{D_k}  \left| \langle \Lambda_i^{(k)} | \Tr_{k+1} \left[ \rho^{(k+1)} - \rho^{(k+1)}_H \right] |\Lambda_i^{(k)} \rangle \right|^\alpha \nonumber \\
    & = \sum_{i=1}^{D_k}  \left| \sum_{j=1}^{D_{k+1}} \left( \lambda_j^{(k+1)} - \frac{1}{D_{k+1}}  \right) \langle \Lambda_i^{(k)} | \Tr_{k+1} \left[ |\Lambda_j^{(k+1)} \rangle \langle \Lambda_j^{(k+1)} | \right] |\Lambda_i^{(k)} \rangle \right|^\alpha \nonumber \\
    & = \sum_{i=1}^{D_k} \left| \sum_{j=1}^{D_{k+1}} \left( \lambda_j^{(k+1)} - \frac{1}{D_{k+1}}  \right) h_{ij} \right|^\alpha
\end{align}
where $\alpha$ appearing in the superscript is the exponent while $\alpha$ appearing in the subscript is the Schatten-index.
Noting that, $\forall i$, $\tilde{h}_{ij}\equiv \frac{D_k}{D_{k+1}} h_{ij}$ is a probability distribution over $j$, we have
\begin{align} 
    \| \rho^{(k)} - \rho^{(k)}_H \|_\alpha^\alpha 
    & = \sum_{i=1}^{D_k} \left| \frac{D_{k+1}}{D_k} \right|^\alpha \left| \sum_{j=1}^{D_{k+1}} \tilde{h}_{ij} \left( \lambda_j^{(k+1)} - \frac{1}{D_{k+1}}  \right)  \right|^\alpha \nonumber \\ 
    & \leq \left| \frac{D_{k+1}}{D_k} \right|^\alpha \sum_{i=1}^{D_k} \sum_{j=1}^{D_{k+1}} \tilde{h}_{ij} \left| \lambda_j^{(k+1)} - \frac{1}{D_{k+1}}  \right|^\alpha,
\end{align}
due to convexity of the function $f(x) = |x|^\alpha$ for $\alpha \geq 1$.
Finally, using $\sum_i h_{ij} = 1$, we obtain
\begin{align}
    \| \rho^{(k)} - \rho^{(k)}_H \|_\alpha^\alpha 
    & \leq \left| \frac{D_{k+1}}{D_k} \right|^{\alpha - 1} \sum_{j=1}^{D_{k+1}} \left| \lambda_j^{(k+1)} - \frac{1}{D_{k+1}}  \right|^\alpha \nonumber \\
    & = \frac{\| \rho_H^{(k)} \|^\alpha_\alpha }{\| \rho_H^{(k+1)} \|^\alpha_\alpha } \| \rho^{(k+1)} - \rho^{(k+1)}_H\|^\alpha_\alpha
\end{align} 
which implies $\Delta^{(k)}_\alpha\leq \Delta^{(k+1)}_\alpha$ $\forall \alpha \geq 1$.
$\blacksquare$


\section{Kicked Ising model and dual-unitarity \label{app:KIM}}

In this section, we elaborate upon the Kicked Ising model (KIM) and explain how  its dynamics, at certain special system parameters, can equivalently be understood as arising from a  quantum circuit which is dual-unitary. Additionally, we explain how the choice of certain initial states and measurement bases result in a transfer matrix in the space-direction which is unitary. 

To begin, the KIM is a Floquet model acting on a 1D chain of $N$ spin-1/2 degrees of freedom, consisting of two alternating steps: time-evolution under  nearest-neighbor Ising interactions and longitudinal fields, followed by time-evolution under a transverse kick. Concretely, the Floquet unitary is given by:
\begin{align}
U_F = e^{-i h \sum_{i=1}^N Y_i} e^{-i (J \sum_{i=1}^{N-1} Z_i Z_{i+1} + g \sum_{i=1}^N Z_i + b_1 Z_1 + b_N Z_N )}.
\end{align}
Above, $X_i, Y_i, Z_i$ represent standard Pauli matrices at site $i$. $J,g,h$ are (dimensionless) strengths of the Ising, longitudinal, and transverse couplings respectively, while $b_1, b_N$ are boundary terms both fixed to be $\pi/4$, introduced solely for technical simplifications. 
We consider the case when $J,h = \pi/4$, while $g$ is arbitrary, and study dynamics under $t \in \mathbb{N}$ applications of the Floquet unitary, $U_F^t$.

The action of $U_F^t$ can be represented by a brickwork quantum circuit which is dual-unitary (in fact, self-dual). 
This is most easily seen using a tensor network diagrammatic notation.
Let us first introduce the following elementary diagrams:
 \begin{equation}
  \vcenter{\hbox{\includegraphics[scale = 0.65]{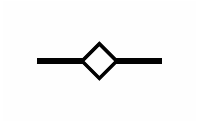}}}    =   \frac{1}{\sqrt{2}}\begin{pmatrix}1 & 1 \\ 1 & -1 \end{pmatrix} , \qquad 
 \vcenter{\hbox{\includegraphics[scale = 0.65]{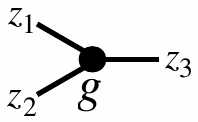}}} = \delta_{z_1 z_2 z_3} e^{-ig(1-2z_1)}.
 \label{eq:tensornetwork1}
\end{equation}
Note a leg $i$ of either diagram carries index $z_i \in \{0,1\}$; we have suppressed writing the indices in the former while explicitly writing  them in the latter.
The former is the standard (unitary) Hadamard gate $H$, while the latter is a three-legged tensor
which has entries $e^{\mp ig}$ if $z_1=z_2=z_3=0 (1)$ and 0 otherwise.

 As is standard with tensor network manipulations, we can contract these tensors with one another or with quantum states. 
We note all equalities presented below are  ``up to (irrelevant) global phases''.
As an example, the contraction of two three-legged tensors yields a four-legged tensor: 
\begin{align}
    \vcenter{\hbox{\includegraphics[scale = 0.65]{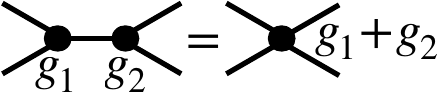}}}. 
    \label{eqn:4leg}
\end{align}
Indeed, one can verify that both the left and right hand side diagrams  are equal to $e^{\mp i(g_1+g_2)}$  if the  indices of all four legs are $0(1)$, and equal to 0 otherwise.

\begin{figure}
    \centering
    \includegraphics[width=0.75\textwidth]{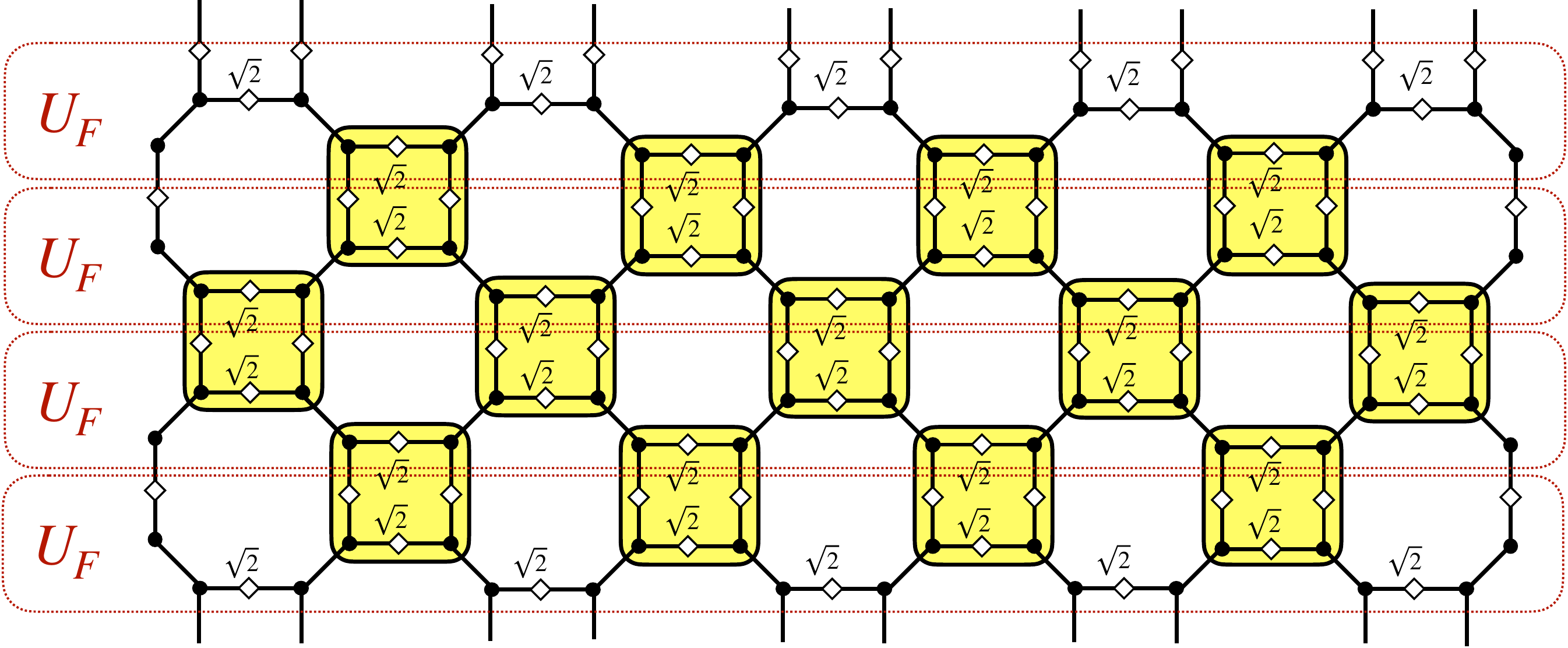}
    \caption{Representation of $U_F^t$ as a brickwork quantum circuit, illustrated here for $N = 10$ and $t = 4$. Each black node implicitly carries a factor $g/2$.
    A single application of the Floquet unitary $U_F$ is given by the action of all the tensors in one red box. 
    Time runs from bottom to top.}
    \label{fig:KIM_Brickwork}
\end{figure}

With this notation, evolution by nearest-neighbor Ising interactions and transverse fields in the $y$-direction, which defines the Floquet unitary $U_F$, can be cast as:
\begin{equation}
e^{-i \frac{\pi}{4} Z \otimes Z} = \sqrt{2} \times \vcenter{\hbox{\includegraphics[scale = 0.65]{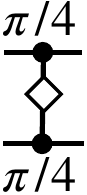}}},   
\qquad e^{-i \frac{\pi}{4} Y} = \vcenter{\hbox{\includegraphics[scale = 0.65]{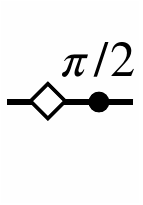}}},
\label{eqn:Ising_int}
\end{equation}
where their action is to be read from right to left.
Finally, the quantum circuit corresponding to $t$ applications of $U_F^t$ can be   drawn as in Fig.~\ref{fig:KIM_Brickwork}.

Referring to Fig.~\ref{fig:KIM_Brickwork}, we see by a judicious arrangement of the tensors that it is a brickwork circuit, made of basic two-qubit quantum gates 
\begin{equation} \vcenter{\hbox{\includegraphics[scale = 0.5]{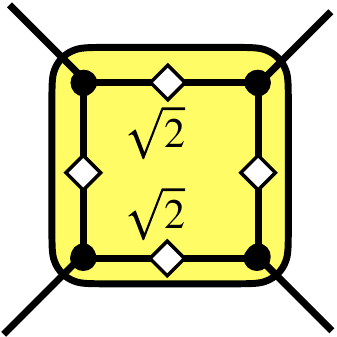}}},
  \end{equation}
  where each black node harbors the factor $g/2$.
Since this tensor represents a unitary operator $U$ when read from bottom to up,  it is evident from symmetry that its space-time dual $\tilde{U}$, defined to be the operator when the diagram is read from right to left, is also unitary. This implies the gate is DU (more precisely, it is self-dual, since $U = \tilde{U}$), and hence dynamics by the KIM at these parameters is equivalent to that of a DU circuit. 
In terms of the parameterization Eq.~\ref{eqn:DU_qubits} of all DU qubit gates stated in the main text $
(r_1 \otimes s_2) \mathsf{SWAP} e^{-iJZ_1Z_2} (u_1 \otimes v_2)$, 
$U$  corresponds to parameters $r_1 = s_2 = e^{-i \frac{g}{2}Z} H$, $u_1 = v_2 = e^{i \frac{\pi}{4} Z} H e^{-i \frac{g}{2}Z} $, and $J = \pi/4$ (up to an overall irrelevant $U(1)$ phase).

\begin{figure}
    \centering
    \includegraphics[width=0.75\textwidth]{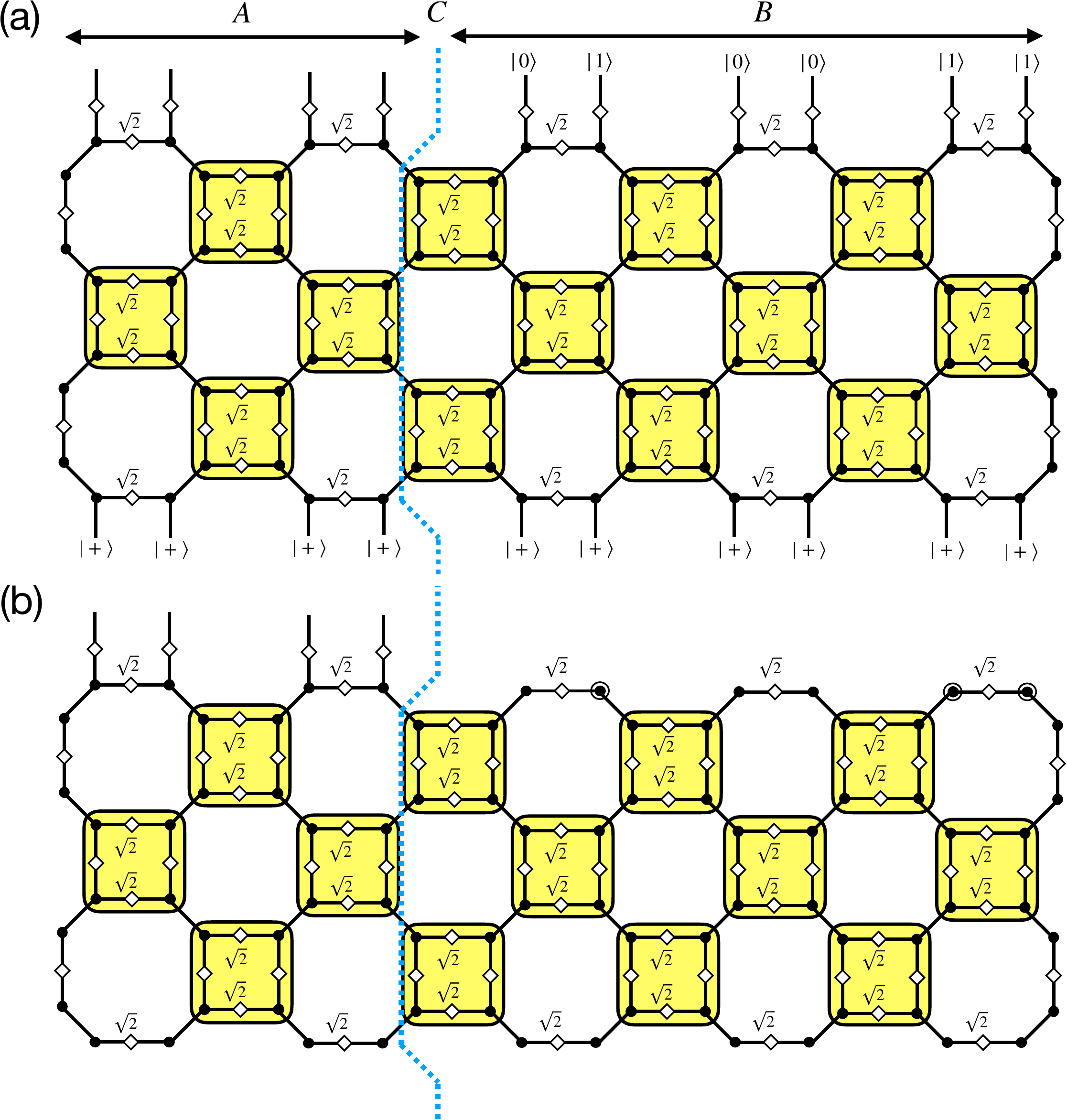}
    \caption{(a) Tensor network representation of unnormalized projected state $(\mathbb{I}_A \otimes \langle \mathbf{z}|) U_F^t |+\rangle^{\otimes N}$ supported on region $A$, illustrated for the measurement outcome $|\mathbf{z}\rangle = |010011\rangle$ on the bath $B$. Each black node carries the value $g/2$.
    (b) Equivalent tensor network diagram in (a), simplified using the relations Eqs.~\eqref{eq:contract1} and \eqref{eq:contract2}.
    This gives an interpretation of the projected state as arising from an instance of a random {\it unitary} quantum circuit evolution in the spatial direction, where the randomness enters as different boundary conditions at the end of the dual chain, indexed by different measurement outcomes.
    }
    \label{fig:KIM_Projected}
\end{figure}

We thus see that the KIM dynamics at parameters $(J,h) = (\pi/4,\pi/4)$ and arbitrary $g$ is DU.
Next, in the study of emergent state designs of the projected ensemble by Ref.~\cite{ho_exact_2022}, initial states consisting of $x$-polarized state $|+\rangle^{\otimes N}$ and measurements in the computational $z$-basis ($z = 0,1$) were considered. 
The tensor network representation of  an (unnormalized) projected state $(\mathbb{I}_A \otimes \langle \mathbf{z}|) U_F^t |+\rangle^{\otimes N}$ can be drawn, as shown in Fig.~\ref{fig:KIM_Projected}(a).
To simplify the figure, the following relations are useful: the contraction  of a three-legged tensor with a local state $|+\rangle = \frac{1}{\sqrt{2}}(|0\rangle + |1\rangle)$  yields
\begin{align}
 \vcenter{\hbox{\includegraphics[scale = 0.65]{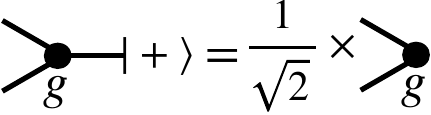}}},
 \label{eq:contract1}
\end{align}
which becomes proportional to  a unitary effecting a $z$-rotation: $\frac{1}{\sqrt{2}}e^{-i g \sigma^z} = \frac{1}{\sqrt{2}} \begin{pmatrix} e^{-i g} & 0 \\ 0 & e^{ig} \end{pmatrix}$;
while a measurement in the computational basis at site $i$ is represented by a contraction with an outcome state $|z_{B,i}\rangle$,  yielding two possibilities:
\begin{align}
\vcenter{\hbox{\includegraphics[scale = 0.65]{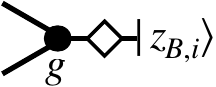}}} =
\begin{cases}
     \vcenter{\hbox{\includegraphics[scale = 0.65]{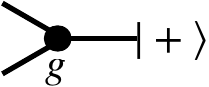}}} &  \text{ if } z_{B,i}~=~0,  \\
          \vcenter{\hbox{\includegraphics[scale = 0.65]{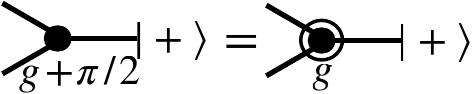}}} & \text{ if } z_{B,i}~=~1.
\end{cases}
\label{eq:contract2}
\end{align}
Note a circled node symbol has been introduced in the last diagram to denote an extra phase angle of $\pi/2$  when the measurement outcome  $z_{B,i}$\,$=$\,$1$. 
Using these, the tensor network representation  of the projected state, Fig.~\ref{fig:KIM_Projected}(a), can be seen to be equivalent to that of Fig.~\ref{fig:KIM_Projected}(b), which describes unitary dynamics in the spatial direction. Thus, the bulk dynamics, initial states and measurement bases considered in Ref.~\cite{ho_exact_2022} are in the DU${}^+$ class.

\section{Review of dynamical purification \label{app:mixedphase}}

Here we provide a brief review of dynamical purification, with a focus on the main result used in this work (the scaling of purification time in the mixed phase). For more thorough discussions of the many aspects of this topic, we refer the reader to Refs.~\cite{gullans_dynamical_2020, gullans_scalable_2020, fidkowski_how_2021, li_statistical_2021, fan_self-organized_2021}.

\subsection{Genereal aspects}
We consider a $1+1$D model of monitored quantum dynamics, in which $N$ qubits evolve under a brickwork circuit of random unitary gates and each qubit is projectively measured with probability $p$ after each gate. The input state is taken to be $\rho_{\rm in} = \mathbb{I}/2^N$ and the dynamics runs to time $T$. 
The question of dynamical purification is: how long must we wait in order for the output state to become (approximately) pure? 
More precisely, there is a family of possible output states $\{\rho_{\mathbf m}(T)\}$, indexed by the measurement record $\mathbf{m}$ (the set of all measurement outcomes obtained during the dynamics up to time $T$); these have an average (von Neumann or Renyi) entropy $\overline{S}(T) = \mathbb{E}_{\mathbf m}[S(\rho_{\mathbf m}(T))]$. Initially, $\overline{S}(0) = N$ bits. The decay of this average entropy defines the {\it purification time} $\tau_p \equiv \min\{ T: \overline{S}(T) \leq \epsilon \}$, for some arbitrary small threshold $\epsilon \ll 1$.
Remarkably, the scaling of $\tau_p$ depends sharply on the rate of measurement $p$: there exists a critical rate $0<p_c < 1$ such that
\begin{equation}
    \tau_p(N) \sim \left\{ 
    \begin{aligned}
    \exp(N) & \text{ if } p<p_c \text{ (mixed phase)}, \\
    {\rm poly}(N) & \text{ if } p=p_c \text{ (critical point)}, \\
    \log(N) & \text{ if } p>p_c \text{ (pure phase)}.
    \end{aligned}
    \right.
\end{equation}
We may interpret $\tau_p$ as a ``memory time''~\cite{fidkowski_how_2021}. Indeed, the entropy $\overline{S}(T)$ is also equal to the average mutual information between the input and output states (e.g., one may view $\rho_{\rm in} = \mathbb{I} / 2^N$ as one half of a maximally-entangled state by introducing a register of $N$ reference qubits); once $\overline{S} \simeq 0$, the output state contains no information about the input state. Thus we may say that information about the input is forgotten over a time scale $\tau_p$.

\begin{figure}
    \centering
    \includegraphics[width=\textwidth]{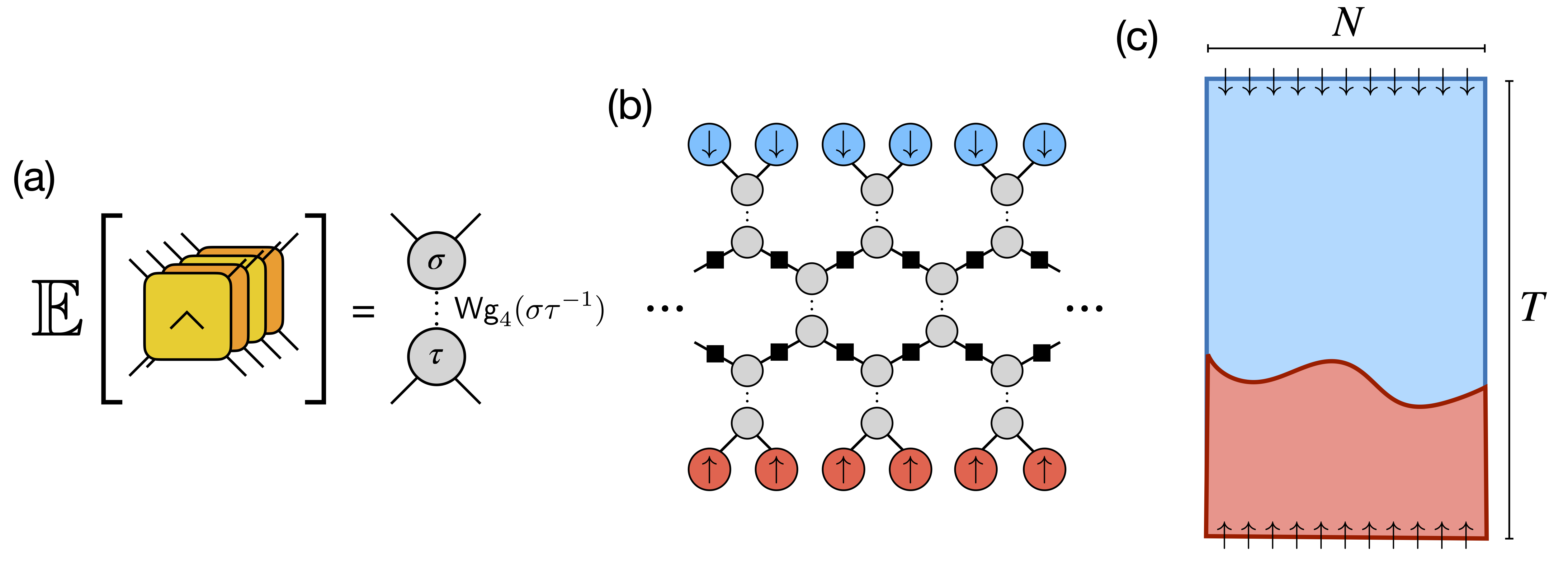}
    \caption{Statistical mechanics of dynamical purification.
    (a) Averaging multiple copies of a Haar-random gate, $(u\otimes u^\ast)^{\otimes n}$ ($n=2$ is shown), yields a sum over permutations $\sigma, \tau \in S_n$, Eq.~\eqref{eq:weingarten}. 
    (b) Averaging all gates in the circuits yields the partition function for a lattice magnet. Circles represent $S_n$-valued ``spins'', squares along the bonds represent the effect of measurements. Boundary conditions at the top and bottom (polarized $\uparrow/\downarrow$) are those relevant to the process of dynamical purification. 
    (c) Coarse-grained description of the entropy $S(T)$ in the mixed phase: the boundary conditions seed $\uparrow$/$\downarrow$-polarized domains in the bulk, which meet at a domain wall. The domain wall costs energy $\sigma N$ and may occur at $\sim T$ positions. (For $T\ll \exp(N)$ the dominant configuration features a single domain wall.) }
    \label{fig:dyn_purification}
\end{figure}

\subsection{Mapping to statistical mechanics}

The existence of the mixed phase, and its stability to a finite rate of measurement for an exponentially long time, is at first glance surprising~\cite{chan_unitary-projective_2019}.
To develop intuition about this phase, it is helpful to introduce a powerful mapping to statistical mechanics which emerges in the case of random circuits~\cite{nahum_quantum_2017, zhou_emergent_2019, zhou_entanglement_2020}. 
Let us denote by $E_{\mathbf m}$ the (non-unitary) linear operator corresponding to the measurement record $\mathbf m$;
the circuit's output states are $\rho_{\mathbf m} \propto E_{\mathbf m} E_{\mathbf m}^\dagger$ and each state occurs with probability $p_{\mathbf m} = \Tr(E_{\mathbf m} E_{\mathbf m}^\dagger) / 2^N$.
Any valid ``order parameter'' for the entanglement phase transition must be a {\it nonlinear} function of the operators $\{ E_{\mathbf m}\otimes E_{\mathbf m}^\dagger \}$ (a linear function $\mathcal{F}$ would, upon averaging over random circuits, become independent of $p$ and insensitive to the transition). 
Let us then take $\mathcal{F}$ to be a homogeneous function of degree $n$ in $E_{\mathbf m} \otimes E_{\mathbf m}^\dagger$. The average over unitary gates drawn from the Haar measure on $U(4)$ can be performed exactly via the Weingarten calculus:
\begin{equation}
    \mathbb{E}_{u\sim \text{Haar}} \prod_{i=1}^n u_{a_i b_i} u^\ast_{c_id_i} = \sum_{\sigma, \tau \in S_n } {\sf Wg}_4(\sigma \tau^{-1}) \prod_{i=1}^n \delta_{a_i, c_{\sigma(i)}} \delta_{b_i,d_{\tau(i)}} \label{eq:weingarten}
\end{equation}
where ${\sf Wg}_d$ are the Weingarten functions for $U(d)$, $\sigma$ and $\tau$ are permutations in the symmetric group $S_n$.
Thus each gate, upon averaging, produces a summation over two permutations in $S_n$, as illustrated in Fig.~\ref{fig:dyn_purification}(a).
Iterating over all gates, we obtain a sum over many $S_n$-valued ``spins'' $\sigma_{x,t}$, $\tau_{x,t}$, each associated to a point $(x,t)$ on a two-dimensional grid (the original $1+1$-dimensional quantum circuit), show in Fig.~\ref{fig:dyn_purification}(b).
This sum is naturally interpreted as the partition function for a two-dimensional, $(n!)$-state lattice magnet. 
The measurement rate in the real circuit affects the coupling between spins in the stat-mech model, and can drive a transition from a paramagnet (pure phase) to a ferromagnet (mixed phase). 

Different order parameters for the measurement-induced entanglement transition map onto the free energy of this magnet under different boundary conditions~\cite{vasseur_entanglement_2019, bao_theory_2020, li_statistical_2021, weinstein_measurement-induced_2022}.
Notably, for the Renyi entropy of a subsystem $A$, one has a boundary condition $e \in S_n$ (identity permutation) at the complementary subsystem $\bar{A}$, corresponding to ``tracing out'', and a boundary condition $\chi \in S_n$ (cyclic permuation) at $A$, corresponding to taking powers of the reduced density matrix.
We focus on the case of $n=2$ replicas for simplicity and denote these two special states as $e = \uparrow$ and $\chi = \downarrow$.
The Reny\'i entropy of $A$ is given by the free-energy cost $\delta F$ of flipping the boundary polarization of the magnet from $\uparrow$ to $\downarrow$ in $A$, while $\bar{A}$ is kept in the $\uparrow$ state. In the ferromagnetic phase, such change of boundary conditions nucleates a domain of $\downarrow$ spins near $A$, surrounded by a domain wall, and thus the free-energy cost is $\delta F\sim |A|$ (length of subsystem $A$); this corresponds to volume-law entanglement entropy.
Conversely in the paramagnetic phase one has $\delta F = O(1)$ and thus area-law entanglement entropy.

The case of interest for us is that of an $N\times T$ strip, representing the monitored dynamics of $N$ qubits for depth $T$, in the ferromagnetic (volume-law) phase.
We aim to understand the dynamical purification of an initially-mixed state. To characterize this process, we look at the entropy of the output state.
This corresponds to the following set of boundary conditions for the magnet: open boundary conditions at $x=0,N$; polarized boundary conditions $\uparrow$ at $t=0$ (mixed initial state); polarized boundary condition $\downarrow$ at $t=T$ (final state whose entropy is being computed)~\cite{li_statistical_2021}. 
This is sketched in Fig.~\ref{fig:dyn_purification}(b).

For $N \ll T \ll \exp(N)$ (i.e., long time, but not exponentially long), the free energy cost is dominated by configurations with a single domain wall between domains of $\uparrow$ and $\downarrow$ spins, sketched in Fig.~\ref{fig:dyn_purification}(c).
Therefore $S(N,T) \approx \sigma N - \ln(T)$, where $\sigma N$ is the energetic contribution ($\sigma$ is a domain-wall line tension) and $-\ln(T)$ is the entropic term term associated to the multiplicity of locations for the domain wall. 
Thus 
\begin{equation}
    S(N,T) \sim \ln(\tau_p/T) 
\end{equation}
where $\tau_p \sim e^{\sigma N}$. 
For $T \gg \tau_p$ domain walls proliferate and the scaling of entropy is expected to cross over to the scaling~\cite{li_statistical_2021}
\begin{equation}
    S(N,T) \sim e^{-2T/\tau_p}; \label{eq:app_Sansatz}
\end{equation}
This shows the exponential scaling of the purification time with $N$. 
While the argument was formulated strictly speaking only for Haar-random circuits, this ``minimal membrane'' picture for entanglement applies also in more general cases~\cite{zhou_entanglement_2020}.

\subsection{Space-time duals of 1D circuits}

Ref.~\cite{ippoliti_fractal_2022} shows that space-time duals of unitary circuits are generically in the volume-law phase, focusing on the ``order of limits'' $t \gg r \gg 1$ (i.e. unitary circuit depth is taken to infinity first). 
For the setup of interest in this work (asymptotic behavior of dynamical purification in space) one cares about the opposite order of limits, $r\gg t \gg 1$ (i.e. spatial size is taken to infinity first).
In this case, it is possible to explicitly map the space-time dual dynamics to that of a conventional monitored circuit and directly show that it indeed belongs to the mixed phase.

\begin{figure}
    \centering
    \includegraphics[width=0.7\textwidth]{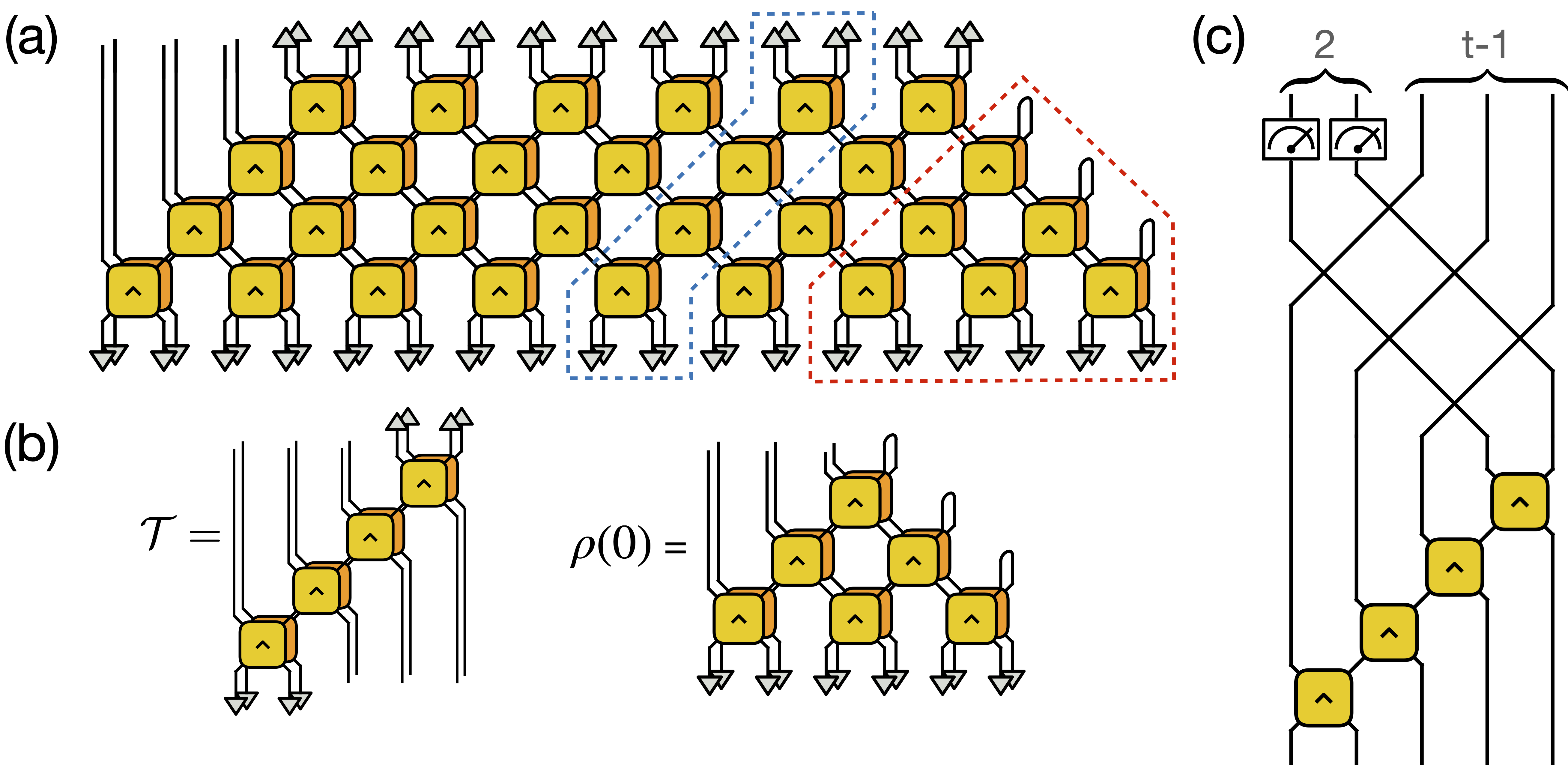}
    \caption{(a) Tensor network diagram for the conditional reduced density matrix $\rho_{\mathbf{z}_1}$ of subsystem $A$ given measurement outcome $\mathbf{z}_1$ on $B_1$ and tracing out $B_2$ (equivalent to Fig.~\ref{fig:purification}(b) but with a single replica instead of two). Depth $t=4$ is shown.
    We may decompose the tensor network into transfer matrices $\mathcal{T}$ along ``light-cone'' cuts, i.e. cuts in the $x=t$ direction (blue dashed box). 
    This leaves out a triangular piece of the tensor network (red dashed box), which can be interpreted as a mixed state $\rho(0)$: $2(t-1)$ qubits in a pure product state evolve under a brickwork circuit, then the $t-1$ qubits on the right are traced out, leaving a $(t-1)$-qubit mixed state $\rho(0)$ on the left.
    (b) The diagrams for $\mathcal{T}$ and $\rho(0)$ shown in isolation.
    (c) Equivalent representation of $\mathcal{T}$ as a monitored circuit on $t+1$ qubits. A layer of a unitary gates in a ``staircase'' pattern couples all $t+1$ qubits; the qubits are then shuffled by $\textsf{SWAP}$ gates (shown as crossing wires) so that the rightmost two end up on the left; these qubits are then measured in the computational basis and reset to $\ket{0}$. Iteration of this circuit $r$ times on the initial state $\rho(0)$ yields the conditional density matrix on $A$ (up to final unitary gates on $A$ beyond the past light cone of $B_1$, omitted here). }
    \label{fig:std_monitored_mapping}
\end{figure}

By segmenting the circuit along ``light-cone cuts'' (i.e. cuts in the $x = t$ direction), as shown in Fig.~\ref{fig:std_monitored_mapping}(a), we may view the spatial evolution as a sequence of transfer matrices $\mathcal{T}$ acting on a mixed initial state $\rho(0)$ of $t-1$ qubits.
The transfer matrix $\mathcal{T}$ consists of the following steps:
\begin{enumerate}[label = (\roman*)]
    \item inject two fresh ancillas in the $\ket{00}$ state on the left,
    \item apply ``staircase'' of unitary gates to all $t+1$ qubits,
    \item projectively measure the two rightmost qubits in the computational basis and discard, obtaining an output state on $t-1$ qubits.
\end{enumerate}
The tensor network diagrams for $\mathcal{T}$ and $\rho(0)$ are separately shown in Fig.~\ref{fig:std_monitored_mapping}(b).
Finally, instead of removing qubits on the right and injecting new ones on the left, we may rewrite $\mathcal{T}$ as an equivalent circuit, shown in Fig.~\ref{fig:std_monitored_mapping}(c), on a fixed system of $t+1$ qubits. A staircase of $\textsf{SWAP}$ gates moves the qubits around so that measured qubits can be recycled on site (analogous to the idea of ``holographic'' tensor networks~\cite{foss-feig_holographic_2021}).

The circuit we obtain contains two measurements for every $t$ entangling gates, giving a measurement rate $p \sim 2/t$.
Thus, the limit of a large ``dual system'' $t\to\infty$ automatically entails a limit of rare measurements $p\to 0$, which guarantees that the monitored dynamics is in the mixed phase, and the entropy of $\rho(0)$ thus decays for larger $r$ as
\begin{equation}
    S(r) \sim e^{-2r/\xi_p(t)}, \qquad \xi_p(t) \sim \exp(t)
\end{equation}
following Eq.~\eqref{eq:app_Sansatz} with $T\leftrightarrow r$, $\tau_p \leftrightarrow \xi_p$.

\section{Existence of a limit state \label{app:limit_state} }

In this Appendix we prove some results that are used towards showing the existence of a limit state $\rho_\infty^{(k)}$ in   Sec.~\ref{sec:du}.

We begin by remarking on the structure of channels $\mathcal{D}_r^\prime$ from Eq.~\eqref{eq:dprime_channels}. First of all, these channels are unitarily equivalent to $\mathcal{D}[\rho] = \frac{1}{4} \sum_\alpha K_\alpha \rho K_\alpha^\dagger$, with $K_{\alpha} = (\sigma_{t}^\alpha)^{\otimes k}$; we focus on the channel $\mathcal{D}$ in the following. 
We may decompose the state $\rho$ into Pauli strings on the $k$-fold replicated spin chain, and note that the Kraus operators $K_\alpha = (\sigma_{t}^\alpha)^{\otimes k}$ (for $\alpha = x, y, z$) are three of these basis operators.
Now, Pauli strings either commute or anticommute with each $K_{\alpha}$.
Let $O K_\alpha = (-1)^{s_\alpha} K_\alpha O$, $s_\alpha \in \mathbb{Z}_2$; by using the fact that $K_x K_y K_z \propto I^{\otimes k}$, we have $s_x \oplus s_y \oplus s_z = 0$ (sum is modulo 2).
Thus we see immediately that the only possible scenarios are $\mathbf{s} = (0,0,0)$ or $\mathbf{s} = (1,1,0)$ and its permutations. 

Thus the basis of Pauli strings can be partitioned into four sets: operators $O_\text{all}$ that commute with {\it all} $K_{\alpha}$'s; operators $O_\text{0,x}$ that commute with $K_0$ and $K_x$ while anticommuting with $K_{y}$ and $K_z$; and similarly-defined operators $O_\text{0,y}$ and $O_\text{0,z}$.
A density matrix can be uniquely decomposed into 
\begin{equation} 
\rho = \rho_\text{all} + \sum_{\alpha=x,y,z} \rho_{0,\alpha} \;.
\end{equation}
It is immediate to verify the following statements (recall $\|A\|_F = \sqrt{\Tr(A^\dagger A)}$ is the Frobenius norm):
\begin{align}
\mathcal{D}[\rho] & = \rho_\text{all} \\
\| \rho \|_F^2 & = \| \rho_\text{all} \|_F^2 + \sum_{\alpha = x,y,z} \left\| \rho_{0, \alpha} \right\|_F^2
\end{align}
From these it follows that
\begin{equation}
\| \rho \|_F^2 - \|\mathcal{D}[\rho] \|_F^2 = \left\| \rho-\mathcal{D}[\rho]  \right\|_F^2 \;.
\label{eq:norm_diff}
\end{equation}

\noindent {\bf Lemma.} {\it Every circuit instance admits a limit state $\rho_\infty^{(k)} = \lim_{N_B\to\infty} \rho^{(k)}$.}

\noindent {\it Proof.} As $\mathcal{D}_r'$ are quantum channels, the sequence of 2-norms $\{n_r\equiv \|\rho^{(k)}_r\|_F: r\in\mathbb{N}\}$ is non-increasing. It is also bounded from below, thus it has a finite limit for $r\to\infty$ (its infimum).
By using Eq.~\eqref{eq:norm_diff}, we have
\begin{equation}
n_r^2 - n_{r+1}^2  =  \| \rho_r^{(k)} - \mathcal{D}_r'[\rho^{(k)}_r] \|_F^2 
= \| \rho_r^{(k)} - \rho_{r+1}^{(k)} \|^2_F \label{eq:norm_diff2}
\end{equation}
Thus the states $\{ \rho^{(k)}_r \}$ form a Cauchy sequence, which converges (since the space of states is metrically complete) to a limit state $\rho_\infty^{(k)}$. $\blacksquare$

\noindent {\bf Corollary.} {\it In every circuit instance, $\| [\rho_r^{(k)}, K_\alpha(r)]\|_F \to 0$ as $r\to\infty$.}

\noindent {\it Proof.}
It follows from Eq.~\eqref{eq:norm_diff2} that $\lim_{r\to\infty} \| (\rho_r^{(k)})_{0, \alpha}\|_F = 0$ for all $\alpha = x, y, z$. 
Since 
\begin{align}
[\rho_r^{(k)}, K_x(r)] 
& =  \left( \rho_r^{(k)} - K_x(r) \rho_r^{(k)} K_x(r) \right) K_x(r) \nonumber \\
& =  \left[ (\rho_r^{(k)})_{0,y} + (\rho_r^{(k)})_{0,z}\right] K_x(r)
\end{align}
(and similarly for the commutators with $K_y$ and $K_z$),
the conclusion follows. $\blacksquare$


\section{Lemmas on universal gate sets \label{app:universal} }

In this Appendix we prove some results used to prove the emergence of exact state designs in Sec.~\ref{sec:du}, having to do with universality of the gate set $G$.

We begin by defining a slightly modified notion of universality for a gate set:

\noindent {\bf Definition.} {\it A set $G \subset U(4)$ of two-qubit gates is \emph{``brickwork-universal''} if the set of brickwork circuits on $N$ qubits, with open boundary conditions, composed of gates in $G$, is dense in the unitary group $U(2^N)$.}

Note that this notion is more restrictive than conventional universality~\cite{barenco_elementary_1995}, as it imposes a restriction on the architecture of the circuit to be used to approximate arbitrary unitaries.
Next, we show that open subsets $G\subset\mathfrak{DU}$ as considered in Sec.~\ref{sec:du} fall within this definition:

\noindent {\bf Lemma.} {\it Any open subset $G \subset \mathfrak{DU}$ is a brickwork-universal gate set.}

\noindent {\bf Proof.} Let us take any gate $U\in G$ and build a brickwork layer $\mathbb{U} = (\mathbb{I} \otimes U^{\otimes N/2 -1} \otimes \mathbb{I}) \cdot U^{\otimes N/2}$ (where $\mathbb{I}$ is the single-qubit identity and we take $N$ to be even). This is a unitary matrix with eigenphases $\{ \phi_n:\ n=1,\dots 2^N\}$. The distance of the time evolution $\mathbb{U}^t$ from the identity is given by $2\sum_n [1-\cos(\phi_n t)]$, a quasiperiodic (or possibly periodic) function. This distance can be made arbitrarily small by choosing a suitable $t = T\in \mathbb N$, according to the theory of Poincar\'e recurrence; thus $\mathbb{U}^T =_\epsilon \mathbb{I}^{\otimes N}$, where $=_\epsilon$ denotes equality up to error $\epsilon$.
By exploiting the fact that $G$ is an open set, we can now make small changes to the parameters of the gate $U$ at one of the bonds that are acted upon in the final time step\footnote{For this bond to be acted on at the final time, $i$ must have a fixed parity.}, say $(i,i+1)$. We obtain a circuit 
$\mathbb{U}' \mathbb{U}^{T-1} =_\epsilon \mathbb{U}' \mathbb{U}^\dagger$
where $\mathbb{U}'$ is the modified final time step. 
Using the parametrization of dual-unitary gates $U = r_1\otimes s_2 \cdot {\sf SWAP} \cdot e^{-i JZ_1 Z_2} \cdot u_1 \otimes v_2$ ($r,s,u,v\in SU(2)$, $J\in \mathbb R$), we see that we may weakly change $r$, $s$ and $J$ (while remaining in the open set $G$) to obtain 
\begin{align}
\mathbb{U}' \mathbb{U}^\dagger 
& = r_i'\otimes s_{i+1}' \cdot e^{-i(J'-J)Z_i Z_{i+1}} \cdot r_i^\dagger \otimes s_{i+1}^\dagger \nonumber \\
& = (r'r^\dagger)_i \otimes (s's^\dagger)_{i+1} \cdot e^{-i (J'-J) (rZr^\dagger)_i (sZs^\dagger)_{i+1}} \;.
\end{align}
This shows that gates in $G$ can generate arbitrary single-qubit rotations on any site\footnote{The gates $r'r^\dagger$, $s's^\dagger$ are small rotations, but by taking sufficiently high powers one can generate any single-qubit rotation.}, as well as entangling operations on half the bonds (recall that the parity of $i$ here is fixed). The same reasoning with a small perturbation at the initial (rather than final) time yields entangling operations on the remaining bonds.
Thus the gate set $G$ is brickwork-universal. $\blacksquare$

Note that, as we did not vary the gates $u$ and $v$, so the assumption that $G$ be an open subset of $\mathfrak{DU}$ can be further tightened.

Circuits built out of brickwork-universal gate sets can approximate any unitary, by definition. In the following, we show that {\it random} circuits built out of such gates in fact approximate every element of the unitary group almost surely, as the circuit depth goes to infinity.

\noindent {\bf Lemma.} {\it Let $B_\epsilon$ be an $\epsilon$-ball in $U(2^N)$. Given a probability distribution $P$ over a brickwork-universal gate set $G$, let $u$ be a brickwork circuit on $N$ qubits of infinite depth (i.e., semi-infinite in the time direction) generated by sampling the gates independently and identically from $P$. Let $u(t)$ be the truncation of $u$ to depth $t$.
Then, the sequence $\{u(t):t\in\mathbb{N}\}$ visits $B_\epsilon$ almost surely.}

\noindent {\bf Proof.} Let $B_\epsilon$ be centered around a unitary $V$. Since the gate set $G$ is brickwork-universal, $V$ can be decomposed, up to accuracy $\epsilon$, into a brickwork circuit of gates in $G$ with finite depth\footnote{Note that this depth scales exponentially in the number of qubits $N$, but here we take $N$ to be a finite constant.} $t_c$ (this is the gate complexity of $V$ relative to $G$).
Thus the probability that $u(t_c) \in B_\epsilon$ is lower-bounded by the probability of randomly sampling the gate decomposition of $V$ (exactly or within $\epsilon$ approximation, depending whether $G$ is discrete or continuous). Let us call this probability $p(V)>0$, and let us define $p_\text{min} = \min_V p(V)$ (note $p_\text{min}>0$ as the unitary group is compact). 
This means that, if we take $T$ as the maximum gate complexity of any element of $U(2^N)$, over a depth $T$ there is a finite probability (bounded below by $p_\text{min}>0$) that the sequence $\{u(t):t=nT+1,\dots (n+1)T\}$ visits $B_\epsilon$, $\forall n$. 
The probability that this doesn't happen over a time $t$ is thus bounded above by $(1-p_\text{min})^{\lfloor t/T \rfloor} \sim e^{-(p_\text{min}/T)t}$.
Thus for $t\to\infty$ the ball $B_\epsilon$ is visited with probability 1. $\blacksquare$
\\


\section{Bounds on the purity of moments \label{app:purity_bound} }

Eq.~\eqref{eq:purity_bound} can be proven by using the convexity of $f(x) = x^k$ for all $k\geq 2$. Let the ensemble be $\mathcal{E} = \{p_i,\ket{i}\}$; then
\begin{align} 
\Tr({\rho^{(k)}}^2)
& = \sum_{i,j} p_i p_j \left| \braket{i}{j} \right|^{2k} 
= \sum_{i,j} p_i p_j f(\left| \braket{i}{j} \right|^2) \nonumber \\
& \geq f \left( \sum_{i,j} p_i p_j \left|\braket{i}{j} \right|^2 \right) 
= \left[ \Tr({\rho^{(1)}}^2) \right]^k \;.
\end{align}

Eq.~\eqref{eq:holder} can be derived by making use of H\"{o}lder's inequality: Let $a,b > 1$ such that $1/a + 1/b = 1$. Then we have
\begin{align}
    \sum_{n} |x_n y_n| \leq \left(\sum_{n}|x_n|^a \right)^{\frac{1}{a}}\left(\sum_{n} |y_n|^b \right)^{\frac{1}{b}},
\end{align}
for $x_n, y_n \in \mathbb{R}$ or $\mathbb{C}$.
To use this, let $x_i = p(i)^{\frac{2}{k}} f(i)$ and $y_i = p(i)^{1 - \frac{2}{k}}$. We choose $a = k$, so that $b = \frac{k}{k-1}$.   Since all quantities are non-negative we can drop the absolute values and obtain
\begin{align}
    \sum_i p(i) f(i) 
    & \leq \left( \sum_i p(i)^2 f(i)^k \right)^\frac{1}{k}
    \left( \sum_i p(i)^\frac{k-2}{k-1} \right)^\frac{k-1}{k} \;.
    \label{eq:applied_holder}
\end{align}
Now, by concavity of $g(x) = x^\frac{1}{k-1}$ ($k\geq 2$) we have 
$\sum_i p(i)^\frac{k-2}{k-1} = \mathbb{E}_i[g(1/p(i))] \leq g[\mathbb{E}_i (1/p(i)) ] = M^\frac{1}{k-1}$. Thus,
\begin{align}
    \sum_i p(i) f(i) 
    & \leq \left( M \sum_i p(i)^2 f(i)^k \right)^\frac{1}{k} \;.
\end{align}
Taking the $k$-th power of both sides yields Eq.~\eqref{eq:holder}.
$\blacksquare$

\twocolumngrid

\bibliography{designs}

\begin{thebibliography}{106}
\bibitem [{Dan()}]{Daniel}%
  \BibitemOpen
  \href@noop {} {}\bibinfo {note} {In a previous version of our manuscript, we
  provided the proof of monotonicity for the Schatten-indices $\alpha = 1,2$
  and $\infty$. We credit Daniel K.~Mark for explaining to us how this result
  can be generalized to all $\alpha \geq 1$.}\BibitemShut {Stop}%
\end{thebibliography}

\end{document}